# Early Life Height and Weight Production Functions with Endogenous Energy and Protein Inputs

Esteban Puentes, Department of Economics, Universidad de Chile[1][*]

Fan Wang, Department of Economics, University of Houston

Jere R. Behrman, Departments of Economics and Sociology and Population Studies Center, University of Pennsylvania.

Flavio Cunha, Department of Economics, Rice University

John Hoddinott, Division of Nutritional Sciences and the Charles H. Dyson School of Applied Economics and Management, Cornell University and International Food Policy Research Institute

John A. Maluccio, Department of Economics, Middlebury College

Linda S. Adair, Department of Nutrition, University of North Carolina

Judith B. Borja, USC-Office of Population Studies Foundation, Inc and Department of Nutrition and Dietetics, University of San Carlos, Cebu, Philippines

Reynaldo Martorell, Rollins School of Public Health, Emory University

Aryeh D. Stein, Rollins School of Public Health, Emory University

February 2016

[1] Corresponding author, email: epuentes@fen.uchile.cl, phone : +56 2 29783455, fax: +56 2 29783413, Address: Diagonal Paraguay 257, of 1501, Zip Code: 8330015.

[*] The authors thank reviewers on previous versions for useful comments and Grand Challenges Canada (Grant 0072-03), Bill & Melinda Gates Foundation (Global Health Grant OPP1032713), and the Eunice Shriver Kennedy National Institute of Child Health and Development (Grant R01 HD070993) for financial support. The funders have no involvement in any part of the research project.

# Abstract

We examine effects of protein and energy intakes on height and weight growth for children between 6 and 24 months old in Guatemala and the Philippines. Using instrumental variables to control for endogeneity and estimating multiple specifications, we find that protein intake plays an important and positive role in height and weight growth in the 6-24 month period. Energy from other macronutrients, however, does not have a robust relation with these two anthropometric measures. Our estimates indicate that in contexts with substantial child malnutrition, increases in protein-rich food intake in the first 24 months can have important growth effects, which previous studies indicate are related significantly to a range of outcomes over the life cycle.

1. Introduction

Inadequate child growth and weight gain are of paramount concern. Approximately 165 million children under five years old in developing countries are stunted and 100 million are underweight (Black et al., (2013)). Growing evidence indicates that early-life undernutrition is associated with, and likely in part causes, reduced education, adult cognitive skills, and wages (Grantham-McGregor, et al., (2007); Engle, et al., (2007); Engle, et al., (2011); Victora, et al., (2008); Hoddinott, et al., (2008, 2013); Behrman, et al., (2009); and Maluccio, et al., (2009)).

Despite widespread concern about early-life undernutrition there is limited systematic knowledge about production technologies for key outcomes, particularly height and weight, needed to inform more-effective program and policy design. This gap is partially due to inherent difficulties in modeling these complex biological and behavioral processes—often strong assumptions are required for estimation, so that it is difficult to make definitive conclusions. A major challenge in estimating production functions for height and weight is that inputs reflect behavioral choices. Using data from the same Philippine study analyzed in this paper, Akin et al. (1992) and Liu, Mroz and Adair (2009) find that families allocate nutrients to compensate for prior poor health. Where allocations reflect compensatory behaviors that are not controlled for in the estimation, the estimated effect of nutrients on growth can be biased.

Another challenge is measurement error in inputs. Using related data from Guatemala, Griffen (2016) finds that estimates of energy effects on height are substantially larger using instrumental variables (IV) than with ordinary least squares (OLS) probably in part due to measurement error.

In this paper, we examine relations between energy intake and: 1) linear growth and 2) weight gain. We use longitudinal data from Guatemala and the Philippines that includes detailed information on anthropometric outcomes, nutrition and other inputs collected at intervals of two-three months to estimate height and weight production functions for children in the critical age range 6-24 months. In our specifications, height and weight depend on lagged height and weight, energy intakes, breastfeeding, diarrhea, and individual fixed endowments. We combine individual fixed-effects (FE) with instrumental variables (IV) to control for both endogeneity and measurement error.

This paper presents three important methodological contributions. First, we estimate production functions for two countries, Guatemala and the Philippines, and for two anthropometric measures, height and weight, which allows us to compare the robustness of our findings across different settings and anthropometric outcomes. Second, we improve on previous IV literature on growth by providing details of instrument selection and an assessment of how the results are robust to changes in the instrument set. We present estimates for numerous instrument combinations, putting emphasis on those judged more reliable based on over-identification and weak instrument tests. Third, in addition to considering total energy intake, which is the nutritional input usually considered in the economics literature, we disaggregate energy intake into two components: proteins and (all) other macronutrients (which we refer to as "non-proteins", meaning fat and carbohydrates). This emphasis on dietary quality, highlighted by Arimond and Ruel (2004), is especially relevant because it may help design interventions that better reduce stunting and underweight. We find robust and positive effects of proteins on height and weight growth. Energy from other macronutrient consumption (non-proteins), is not systematically related to these anthropometric measures, which suggests that protein-rich foods are particularly important for growth of undernourished children.

2. **Specifications of Height and Weight Production Functions and Identification**

2.1. Input Selection

Our choice of inputs is guided by Black et al. (2008) who argue that inadequate diet and disease are the main immediate causes of stunting and wasting. With respect to diet, two energy sources have been identified as being especially important for child growth: proteins and non-protein energy from other macronutrients. Infants require certain minimum amounts of energy and proteins to maintain long-term good health but these requirements are heterogeneous and depend on several factors including weight and whether the child is breastfed (FAO (2001); WHO (2007)). Children's energy requirements are partly driven by energy costs of linear growth, which has two components: 1) energy needed to synthesize growing tissues and 2) energy stored in these tissues (FAO, (2001)). These comprise approximately one-third of total energy requirements during the first three months of life, but despite increasing in absolute terms they decline to only 3% by age 24 months, in part because overall energy requirements increase substantially with body size. Proteins are needed to balance nitrogen loss, maintain the body's muscle mass, and fulfill needs related to tissue deposition (WHO, (2007)). There is also evidence from research on animals that protein provides anabolic drive for linear bone growth (WHO, (2007)).[2]

To study the relative importance of protein and non-protein sources, we first examine the relationship between total energy and height and weight and then consider the potential for separate roles of the two at once in a single growth model. The comparison of proteins with non-proteins highlights the relative importance of proteins in children's diets and informs what types of interventions might have greater impact on height and weight.[3] There is a limited literature focused on the distinction between total energy and protein energy. Pucilowska et al. (1993) find that high-protein supplementation in Bangladeshi children with shigellosis, a severe bacterial disease, increased weight compared to normal protein diets. A randomized evaluation for children up to 2 years of age in several European countries demonstrated that receiving baby formula with high protein content (% calories from protein) increased weight, but not height (Koletzko et al., (2009)). Both of these study populations, however, are different from the ones we examine. The Bangladeshi sample is restricted to children recovering from shigellosis while the European sample had not experienced the same nutritional deficiencies found in our samples. Using a sample more similar to ours, Moradi (2010) finds that access to high-quality protein, such as from livestock farming, better predicts height in some African countries than other energy sources. Similarly, Baten and Blum (2014), using global information for the first part of the twentieth century, that includes Guatemala and the Philippines, also find that local availability of cattle, milk and meat were an important predictor of adult height.[4]

---

[2] Micronutrients also play important roles in tissue building (WHO, (2007)), but there is limited information about them in our data. Hence our focus on protein and non-protein energy.

[3] While other individual macronutrients may have different relationships with growth (WHO, (2007)), separating them into their components while still treating them as endogenous was empirically infeasible.

A related issue is protein quality. Proteins are composed of amino acids with specific cell functions, and amino acid content defines protein quality. For instance, plant-based proteins lack essential amino acids unlike animal-based proteins (Dewey, (2013)). In addition, plant-based diets have high levels of phytic acid, which might inhibit zinc absorption (Gibson, (2006)), and zinc plays a key role in cellular growth and differentiation (Imdad and Bhutta, (2011)). For animal-based protein, Mølgaard et al., (2011) argue that dairy intake has positive impacts on child growth. Although the mechanism is not entirely clear, this may be due to the stimulating effect on plasma insulin-like growth factor (IGF-1) (Michaelsen, (2013)).

Breastfeeding is another critically important source of nutrition in early life (Black et al. (2013)). In this paper, we have data on breastfeeding status but not on the amount of breast milk consumed. Thus, our energy intake measures exclude energy from breastmilk requiring us to control for breastfeeding status in the models.

---

[4] Relatedly, and using the same data from the Philippines that we use, Bhargava (2016) studies the association of macronutrients (proteins) and micronutrients (calcium) with anthropometrics, finding that both, protein and calcium are strongly associated with height and weight in the first 24 months of life and also on adolescence. However, Bhargava (2016) only controls for individual effects, assuming several time varying variables as exogenous.

Among diseases that affect growth, Walker et al. (2011) suggest that persistent diarrhea and other diseases can have long-lasting effects on children's physical development. Therefore, in our analyses, we incorporate diarrhea as an input, as it is considered a major contributor to stunting, wasting and child mortality (Black et al (2013)).

### 2.2. Height and Weight Production Functions

The main challenges for estimating height and weight production functions include the endogeneity of inputs and measurement error (Behrman and Deolalikar (1988)). To overcome these, we follow the general approach developed in recent research on production function estimation for cognitive and non-cognitive skills (Todd and Wolpin (2003, 2007); Cunha and Heckman (2007)).

Let $h_{i,t}$ denote child $i$ height at age $t$, $w_{i,t}$ weight at age $t$ and $x_{i,j}$ the input (e.g., proteins, non-proteins, or disease) at age $j$. (For simplicity, we present the model with a single input but generalization to several inputs is straightforward.) Fairly general height and weight production functions are:

$$h_{i,t} = \alpha \mu_i + \sum_{j=1}^{t} \beta_{t-j} x_{i,j} + \epsilon_{i,t}^h \quad (1)$$

$$w_{i,t} = \sigma \mu_i + \sum_{j=1}^{t} \delta_{t-j} x_{i,j} + \epsilon_{i,t}^w \quad (2)$$

Where $\mu_i$ is an individual fixed effect (including genetic endowments and fixed parental and household characteristics) and $\epsilon_{i,t}^h$ and $\epsilon_{i,t}^w$ are error terms. This formulation allows the entire input history to enter into both equations up to time $t$. Furthermore, it allows for impacts of past inputs on current height and weight and for the possibility that such impacts differ by age. This approach also distinguishes our work from other studies using the same data. Griffen (2016) relies on the fairly strong assumption that past inputs have constant effects on height in Guatemala, so that history plays little role in growth.

Similarly, height production functions estimated by de Cao (2015) in the Philippines, assume that height growth depends only on current inputs.

Because they include individual fixed effects and the entire input history, equations (1) and (2) are difficult to estimate. For example, if inputs are treated as endogenous and an IV approach were used, it would be necessary to have at least one instrument for each period in the entire input history. Thus, instead of directly estimating these two equations, we make two further assumptions that allow less demanding specifications in terms of data and instrument requirements, while remaining more flexible than previous specifications in the literature.

*Assumption 1*. Effects of past inputs follow a monotonic (likely decreasing) pattern at a constant rate $\gamma$ for each period.[5] That is: $\beta_{t-j} = \gamma \beta_{t-1-j}$ and $\delta_{t-j} = \gamma \delta_{t-1-j}$.

*Assumption 2*. The coefficients on inputs in the height function are the same as those in the weight function, up to a multiplicative constant $\delta_{t-1-j} = \frac{1+\sigma}{\alpha} \beta_{t-1-j}$.

Together, these assumptions reduce the set of endogenous variables to a tractable number, thereby reducing the number of required instrumental variables.

From equation (1) and taking first-differences in height we obtain:

$$\Delta h_{i,t} = \beta_0 x_{i,t} + \sum_{j=1}^{t-1} (\beta_{t-j} - \beta_{t-1-j}) x_{i,j} + \epsilon_{i,t}^h - \epsilon_{i,t-1}^h$$

Incorporating the first assumption that $\beta_{t-j} = \gamma \beta_{t-1-j}$, we obtain:

$$\Delta h_{i,t} = \beta_0 x_{i,t} + (\gamma - 1) \sum_{j=1}^{t-1} \beta_{t-1-j} x_{i,j} + \epsilon_{i,t}^h - \epsilon_{i,t-1}^h$$

Next, consider the difference in equations (1) and (2) (after cross multiplication with σ and α):

---

[5] While it seems most likely that nutritional inputs would have a larger impact during the 6-24 month age window we model, assuming it is decreasing is not strictly necessary. The rate can be different for the height and weight equations; we assume that is similar only for illustration purposes.

$$\alpha w_{i,t-1} - \sigma h_{i,t-1} = \sum_{j=1}^{t-1}(\alpha\delta_{t-1-j} - \sigma\beta_{t-1-j})x_{i,j} + \alpha\epsilon_{i,t-1}^{w} - \sigma\epsilon_{i,t-1}^{h}$$

Under the second assumption that $\delta_{t-1-j} = \frac{1+\sigma}{\alpha}\beta_{t-1-j}$, we have:

$$\alpha w_{i,t-1} - \sigma h_{i,t-1} + \sigma\epsilon_{i,t-1}^{h} - \alpha\epsilon_{i,t-1}^{w} = \sum_{j=1}^{t-1}\beta_{t-1-j}x_{i,j}$$

Consequently,

$$\Delta h_{i,t} = \beta_0 x_{i,t} + \alpha(\gamma - 1)w_{i,t-1} - \sigma(\gamma - 1)h_{i,t-1} + \omega_{i,t}^{\Delta h} \quad (3)$$

where $\omega_{i,t}^{\Delta h} = \epsilon_{i,t}^{h} + (\sigma(\gamma - 1) - 1)\epsilon_{i,t-1}^{h} - \alpha(\gamma - 1)\epsilon_{i,t-1}^{w}$.

Under these assumptions, height growth can be expressed as a function of current inputs, past height and weight, and an error involving current ($t$) and previous period ($t$-$1$) shocks. Current inputs enter directly; the full history of past inputs enter indirectly through the lagged height and weight.

We proceed in similar fashion for weight and obtain:

$$\Delta w_{i,t} = \delta_0 x_{i,t} + (\gamma - 1)(1 + \sigma)w_{i,t-1} - \frac{\sigma(\gamma-1)(1+\sigma)}{\alpha}h_{i,t-1} + \omega_{i,t}^{\Delta w} \quad (4)$$

where $\omega_{i,t}^{\Delta w} = \epsilon_{i,t}^{w} + \frac{\sigma(\gamma-1)(1+\sigma)}{\alpha}\epsilon_{i,t-1}^{h} - [(\gamma - 1)(1 + \sigma) + 1]\epsilon_{i,t-1}^{w}$.

As with the change-in-height equation (3), the change-in-weight equation (4) depends on current inputs, past height and weight, and an error including current and previous period shocks.[6]

This framework forms the core of our approach to estimating production functions for height and weight. Estimation of equations (3) and (4) allow recovering $\beta_0$ from equation (1) and $\delta_0$ from (2).

---

[6] Specifications of the change-in-height equation that exclude lagged weight, and the change-in-weight equation that exclude lagged height were also estimated. Results were similar to the more general specification (available on request).

2.3. Estimation and Identification

Although differencing removes individual-level fixed effects and thus controls for important sources of potential bias (unobserved persistent heterogeneity including, e.g., genetic endowments and fixed parental and household characteristics), to consistently estimate the parameters in the relations for change in height (equation (3)) and change in weight (equation (4)), we still need to overcome several endogeneity problems. First, by construction previous height and weight are correlated with the error terms of equations 3 and 4 (see equations (1) and (2)). Moreover, if we assume that the household responds to past shocks as is likely and for which there is evidence for the Philippines (Akin et al. (1992); Liu et al. (2009)), current inputs may be correlated with the error terms.

We address potential endogeneity by using IV, which also addresses bias due to random measurement error in $x$ under the assumption that the instruments are uncorrelated with that measurement error. The set of candidate instruments we use differs by country but draws on plausibly exogenous factors including a randomized intervention in Guatemala and prices of common foods in both countries. We treat market prices as exogenous to households (as in Liu et al. (2009)). Using prices as instruments for inputs is a well-established approach in the estimation of production functions (Todd and Wolpin (2003)). We also include past height and weight measures, $h_{i,t-2}$ and $w_{i,t-2}$ as instruments to help identify the effects of lagged height and weight. (Instruments are described in further detail in Section 3.3.)

Using the available instruments, we endogenize protein and non-protein intakes, as well as lagged height and weight. However, we do not have access to instruments in both countries that also would allow us to control for the potential endogeneity of breastfeeding or diarrhea.[7] Controlling for

---

[7] Previous work using the Philippine data has used rainfall as an instrument for diarrhea (Akin et al., (1992)). We attempted to endogenize diarrhea using spatial and temporal variation in rainfall and

individual-level fixed effects is an important aspect of our approach, however, and goes part way toward addressing their potential endogeneity. For example, fixed effects control for the possibility that certain children have a pre-disposition for diarrhea, or live in particularly unsanitary households. However, if households change breastfeeding practices when health shocks affect their children's health or change sanitary conditions to reduce the diarrhea prevalence, the estimated effects of breastfeeding and diarrhea could be downward-biased. For instance, households that have increased breastfeeding could be compensating for negative health shocks, suggesting a negative relationship between growth and breastfeeding, while correcting for endogeneity could show a positive relationship (and similarly for diarrhea). Because our principal objective is to study the roles of proteins and non-proteins in the production functions, however, we do not emphasize the coefficients for diarrhea and breastfeeding but instead make clear the assumptions under which our primary coefficients of interest are consistently estimated *even if* breastfeeding or diarrhea are endogenous in the model. Our estimation approach is consistent provided the instruments are not correlated with the error term in the production function, conditional on breastfeeding and diarrhea as well as other covariates mention bellow. This is plausible for the same reason that the instruments are exogenous in relation to the energy inputs, e.g., that they are not correlated with individual-level time-varying health shocks.[8]

In principle, there also could be interactions among inputs in the production function, such as between nutrient intakes and diarrhea, or between breastfeeding and other nutrient intakes but a specification incorporating such interactions would be even more challenging to estimate, requiring additional instruments . Given that there are already four variables that we treat as endogenous in our main models (protein, non-protein, lagged height, and lagged weight), we do not estimate models with

---

temperature as instruments in Guatemala, but they had minimal predictive power. To keep the structure parallel across the countries, we do not use rainfall to endogenize diarrhea in either country.

[8] For instance, if some other disease is important in the production function, and we are not including it, our results hold if the instrumental variables are orthogonal to this other disease.

such potential interactions; instead, we studied possible interactions by splitting the sample. For instance, to examine whether diarrhea or breastfeeding interacts with diets, we estimated specifications for the sample that is breastfed and compare the results with the sample that is not breastfed. We carried out a similar exercise for diarrhea. Our results indicate that coefficients are not affected when we separate the sample by breastfeeding types. For diarrhea, there was some evidence of interaction effects, where diarrhea lowers the effects of macronutrients, but because most of the specifications suffer from problems of weak instruments, we are unable to draw strong conclusions.

The estimation of the growth equations also includes an indicator for whether the child was female, number of days since the previous measurement, and age and age squared at time *t*.

Our methods permit us to improve upon the previous literature that investigates the effects of total energy on anthropometrics. Since we do not have a single set of preferred instruments, we are able to robustly study effects of total energy on height and weight across two settings. We do this estimating the changes in height and weight, first using total energy intakes and then separating protein and energy from other macronutrient intakes to examine their relative partial effects in each model.

The final estimating equations for the change in each anthropometric measure $A_{i,t}$ that we estimate, adding the additional controls to equations (3) and (4), are:

$$\Delta A_{i,t} = \lambda^A_{energy} E_{i,t} + \rho^A_1 w_{i,t-1} + \rho^A_2 h_{i,t-1} + \rho^A_3 days\_no\_diar_{i,t}$$

$$+\rho^A_4 bf_{i,t} + \rho^A_5 age_{i,t} + \rho^A_6 age^2_{i,t} + \rho^A_7 female_{i,t} + \rho^A_8 gap\_msmt_{i,t} + \eta^{\Delta A}_{i,t} \quad (5)$$

and

$$\Delta A_{i,t} = \lambda^A_{prot} Prot_{i,t} + \lambda^A_{non\_prot} Non\_Prot_{i,t} + \delta^A_1 w_{i,t-1} + \delta^A_2 h_{i,t-1}$$

$$+\delta^A_3 days\_no\_diar_{i,t} + \delta^A_4 bf_{i,t} + \delta^A_5 age_{i,t} + \delta^A_6 age^2_{i,t} + \delta^A_7 female_{i,t} + \delta^A_8 gap\_msmt_{i,t} +$$

$$v^{\Delta A}_{i,t} \quad (6)$$

where $A_{i,t}$ is either weight $(w_{i,t})$ or height $(h_{i,t})$ of child i at age t; $E_{i,t}, Prot_{i,t}, Non\_Prot_{i,t}$ correspond to the total energy intake, protein intake and non-protein intake; days_no_diar$_{i,t}$ is the number of days without diarrhea between measurements; bf$_{i,t}$ is a dummy variable equal 1 if the child was breastfed during the period leading up to age t; age$_{i,t}$, and age$^2_{i,t}$ are age and age squared; female$_{i,t}$ is a dummy variable equal to 1 if the child is a female; and $gap\_msmt_{i,t}$ is the number of days between measurements. Finally, the error terms in equations (3) and (4) exhibit serial correlation of order one by construction. We use cluster standard errors at the individual level to take into account this serial correlation, and also any possible correlation of individual error terms; using cluster standard errors is more general than a correction for serial correlation. Additionally the error terms are correlated between equations so there are possible efficiency gains of estimating a system of equations. Nonetheless given the already complex nature of the estimation, we estimate single equations. The cluster errors we calculate, therefore, can be seen as an upper bound of the standard errors.

## 3. Data

Estimation of (5) and (6) requires high-frequency longitudinal data in early life that contain information on the outcomes (height[9] and weight) and inputs (proteins and other macronutrients, breastfeeding, and diarrhea), as well as plausibly exogenous instruments. We now describe the data and contexts for two unique studies that fulfill these substantial requirements relatively well, one in Guatemala from the 1970s and the other in the Philippines from the 1980s.

### 3.1. Guatemala

---

[9] In both settings, children under 24 months were measured lying down, per standard anthropometric measurement practice. This measurement is sometimes referred to as length, rather than height.

We use data from The Institute of Nutrition of Central America and Panama (INCAP) 1969-1977 nutritional supplementation trial. Four rural villages from eastern Guatemala were selected, one relatively large pair (~900 residents) and one smaller pair (~500 residents). At the outset, the villages were similar in terms of child nutritional status, measured as height at age three years, and were highly malnourished with over 50% of children severely stunted, i.e., with height-for-age z-score < -3. One large and one small village were randomly selected to receive a high-protein supplement (Atole); the others received an alternative supplement devoid of protein (Fresco). A 180 ml serving of Atole contained 11.5 grams of protein and 163 kcal. Fresco had no protein and a 180 ml serving had 59 kcal. The main hypothesis was that increased protein would accelerate mental development; additionally, it was expected that the high-protein nutritional supplement would affect physical growth. The nutritional supplements were distributed in centrally-located feeding centers in each village (Habicht, Martorell and Rivera, (1995)). Virtually all (>98%) families participated (Martorell et al., (1995)).

From 1969 to 1977, anthropometric measures (height and weight) were taken every three months for all children 24 months of age or under (including newborns entering the study) in the four villages. This yields a maximum usable sample for our analyses of 878 children measured at least twice by the age of 24 months. The amount of supplement intake was recorded daily in all villages. Home dietary information was collected every three months, including the types and amounts (except for breastmilk) of all foods and liquids consumed. These dietary histories were based on a 24-hour recall period in the larger villages and a 72-hour period in the smaller villages (from which we construct daily averages), and permit calculation of protein and non-protein intakes for the 24-hour period by summing the nutritional content for each food item. The survey recorded the total months a child was breastfed. Nutrients from breastfeeding were not included in the nutritional intake calculations. Retrospective information on illness, specifically the length in days of episodes of diarrhea and fever, was collected semi-monthly.

### 3.2. The Philippines

We use the Cebu Longitudinal Health and Nutritional Survey, a survey of Filipino children born between May 1983 and April 1984 in 33 rural and urban communities (barangays) in Metropolitan Cebu. The baseline survey included 3,327 women sampled at a median of 30 weeks of gestation, and yielded a sample of 3,080 singleton live births. This sample also exhibits high levels of undernutrition; at age 24 months, 62% of the children were stunted and 32% underweight. During the first two years of each child's life, data were collected every two months. This included anthropometric measurements, 24-hour dietary recall of types and amounts (except breast milk) of all foods and liquids eaten, breastfeeding, and recent illness history. For breastfed children, the survey also collected the frequency and length of time spent breastfeeding. Total protein and energy intakes were calculated from foods consumed the previous day (24-hour recall method). At each survey, mothers reported whether the child had diarrhea in the past 24 hours, and if so, when the episode began, and the number of days the child had diarrhea during the previous week (Adair et al. 2011). The maximum usable sample of children between 6 and 24 months of age for the Philippines is 2,713.

### 3.3. Variable construction

Linear growth and weight gain are calculated as the difference between consecutive measurements. Although measurements were scheduled at specified intervals (every three months in Guatemala, every two in the Philippines), there were deviations including instances where a scheduled measurement did not occur. Because children experience high growth and growth spurts during the first two years of life, even differences of several days can be associated with significant differences in growth. We account for this by controlling for the exact number of days between measurements.

Ideal data for this analysis would have information on protein and non-protein intakes over the *entire* period between measurements, but even in these uniquely comprehensive studies such detailed information is not available. Therefore, we approximate intakes over the entire period by using the average of the 24-hour intakes calculated from the dietary recall information at the beginning and end of each period (which decreases measurement error relative to using only one point in time) multiplied by the exact number of days between measurements. For Guatemala, we add to this figure the intakes from the supplement (which were measured daily throughout the period) to obtain total protein and other intakes (as well as their sum, measured as total energy).[10]

For breastfeeding, we create a dummy indicator for whether the child was breastfed in the month previous to measurement at time *t*. While this does not fully exploit the detailed information available for the Philippines, it is done to have similar specifications across countries.

The final input we include is diarrhea. For Guatemala, the protocol was to collect information every 15 days, so it is possible to construct the number of days experiencing diarrhea for the complete periods between anthropometric measurements.[11] For the Philippines, it is only possible to construct the number of days with diarrhea during the week previous to each bimonthly anthropometric measurement. To extrapolate this to the full period between measurements, we estimate a count model for number of days with diarrhea for each two-month period with the Guatemalan data and use the estimated parameters from that model to predict number of days each Filipino child had diarrhea in each two-month period.[12]

As outlined in Section 2.3, in our main specifications we instrument for protein, other macronutrient intakes, and lagged height and weight. We now describe in detail the other instruments besides twice lagged height and weight.

---

[10] For Guatemala we use an individual-level fixed-effects model to impute nutrient intakes for approximately 5% of missing observations. See Data Appendix Section 1.

[11] Approximately 45% of such 15-day visits were missed. In those instances, we assume the child had similar diarrhea patterns across all 15-day intervals during that growth period and scale-up the observed number of days accordingly.

[12] See Data Appendix Section 2 for details of the estimation of the count model for diarrhea.

In both countries we use unit prices for various food items, selected with emphasis on foods with high protein content and/or important in the local diet. For Guatemala, prices are averages of national-level prices measured during December each year. We use lagged prices of eggs, chicken, pork, beef, dry beans, corn, and rice. Unit price variables for Guatemala are deflated and measured over the eight-year study period. For the Philippines, we use community-specific prices collected as part of the broader study. Between January 1983 and May 1986, enumerators visited two stores in each community, every other month, and collected prices (and quantity units) for a list of items. Not all items, however, were sold at each store at each visit. Consequently, there is not a complete set of prices for each item from each store (or even from each community in instances where no price was available from either store) in each measurement period. We selected as instruments the prices of dried fish, eggs, corn and tomatoes since these are the ones with the highest frequency in the sample.[13] We use both current and lagged prices of those selected food items. By estimating a large set of instrument combinations, our approach does not depend on any one particular price, avoiding subjective instrument selection.

For Guatemala, we also exploit the experimental variation resulting from the randomized allocation. We use a dummy variable that indicates whether the village had a feeding center that provided the high-protein supplement. We also interact this indicator with the distance from the home of the child to that feeding center. While the presence of a randomized allocation of a high-protein supplement provides an important source of exogenous variation, since there are four endogenous variables, additional instrumental variables also are used, i.e., twice lagged anthropometrics and food prices. For the Philippines we rely on price variation, which, unlike the annual Guatemalan food price data, varies both within-years and spatially, with information on these food items for the majority of measurement periods and each of the 33 communities.

---

[13] See Data Appendix Section 3 for further details on prices.

3.4. Descriptive statistics

Over the period from ages 6 to 24 months, each Guatemalan child is observed an average of 4.3 times and each Filipino child 9.1 times. The sample we describe includes all observations (measurements of children at different ages) with complete information for the following variables: change in height between consecutive measurement periods (linear growth), change in weight between consecutive periods (weight gain), total energy, energy from protein, energy from non-protein, breastfeeding indicator, and days with diarrhea.[14] The final number of observations used in each specification varies depending on the availability of the instrumental variables used in that specification, since instruments for some observations are missing.

Table 1 compares the main variables for both samples. On average and at all ages, the Filipino children in the early 1980s were taller than the Guatemalan children in the 1970s. For example, at 12 months of age, Filipino children were on average 70.7 cm tall, while their Guatemalan counterparts were 1.8 cm shorter. In terms of average weight, however, there were no significant differences between countries—at 24 months, children from both countries averaged 9.8 kg. 44% of the Guatemalan children were stunted, and 27% underweight. The corresponding levels were lower, 25% and 11%, for Filipino children. In 2011 for low- and middle-income countries, average levels of stunting were 28% and of underweight 17%, and 36% and 18% in Africa (Black et al., (2013)). With broadly similar levels of stunting and underweight, thus, our historical samples remain relevant to understanding undernutrition in many countries and regions.

---

[14] For the Philippines, the number of available observations is constant across variables, but decreases with child age due to attrition. For Guatemala, the number of children with available information on intakes and diarrhea is smaller than the number with anthropometric measures because the dietary and morbidity information for infants under 12 months was not collected until 1973.

Table 2 shows that Guatemalan children appear more likely to have been breastfed at all ages. In both countries, breastfeeding declines with age. At six months, 99% of Guatemalan children were breastfed, while at 24 months only 18% were; the proportions were 76% and 14% for Filipino children.

Patterns between diarrhea and age are less clear. In Guatemala, average number of days with diarrhea (per 3-month measurement period) increases with age to 15 months, after which it declines. Levels are relatively lower in the Philippines, fluctuating between about 2 and 6 days (per 2-month period), with no clear age pattern.

For Guatemala, information is complete on all of the instruments except the distance to the feeding center, which is missing for ~5% of observations. For the Philippines, on the other hand, incomplete price availability leads to larger reductions in the sample size. The potential sample has 24,820 child-age observations; the lagged price of corn, which is the most complete, has 18,710 observations and the lagged price of tomatoes, the least complete, has 16,084 observations.

4. Results

4.1. Overview

We estimate height and production functions for children 6 to 24 months, the period widely considered to be a critical window for post-birth nutritional investment.[15] We use Generalized Method of Moments (GMM) for exactly-identified models and Limited Information Maximum Likelihood (LIML) for over-identified models because the latter allows for smaller finite-sample bias (Stock and Yogo, (2005)). As noted, we cluster error terms at the individual level to take into account correlation of individual error terms and serial correlation (Baum et al. (2007)).[16] We first estimate height and weight production functions using only total energy (i.e., the sum of calories from protein and other sources), then we analyze separately the roles of proteins and non-proteins. In all specifications, the energy intakes, lagged height, and lagged weight are treated as endogenous, and we control for breastfeeding, number of days without diarrhea since the previous measurement, child sex, number of days since the previous measurement, and age and age squared.

---

[15] There are additional substantive, as well as practical, reasons for the 6-24 month window. First, during the first six months most infants are breastfed; indeed WHO recommends exclusive breastfeeding from birth to age six months. Therefore, before that age proteins and non-proteins in the diet reflect non-exclusive breastfeeding that could be detrimental to growth. Second, it is not possible to study the production function at earlier ages because our final specification models growth and the candidate instrumental variables include second lags of height and weight (Section 2.2). Because we model growth and use these second lags, however, the analysis does incorporate information on individuals prior to six months of age. Third, while the frequency of measurements differs, both samples have measurements at ages six and 24 months, facilitating comparability.

Because there are many potential instrument combinations, to establish general results that do not depend on one specific instrument combination, we estimated large subsets of all possible combinations. For Guatemala we first restricted the instrument sets to combinations that always had the Atole experiment indicator. Then, we systematically varied inclusion of distance interactions with Atole indicator, second lags of height, second lags of weight, and from two to four of the seven food prices (eggs, chicken, pork, beef, rice, beans and corn). For the Philippines, we systematically varied inclusion of second lags of height, second lags of weight, and from two to six of the eight (four current and four lagged) food prices (eggs, fish, tomatoes and corn). A summary of our instrument combinations is found in the Data Appendix Section 6. For Guatemala, there are 546 specifications (i.e., each with a different instrument set) for the version of the model with total energy (equation (5)) and 525 when proteins and non-proteins are included separately (equation (6)).[17] The total number of specifications estimated for the Philippines is 602 for both models.

---

[16] The specifications also include predicted days of diarrhea. We do not explicitly account this in calculating the standard errors, instead relying on the general correction provided by clustered standard error calculations.

[17] The reduction in specifications arises because 21 specifications that include both protein and non-protein are exactly-identified with three instruments.

For each specification, we calculate the robust versions of the Hansen-J (HJ) over-identification test, the Anderson-Rubin under-identification test (Anderson and Rubin, (1949)), and the Wald F-statistic (robust Cragg-Donald or CD statistic) to detect weak instruments. Since our main models have four endogenous variables and we estimate them assuming heterokedasticity, it is not possible to compare CD statistics with critical values from Stock and Yogo (2005). The robust versions of these tests were developed in Kleibergen and Paap (2006). We also calculate for each endogenous variable Angrist and Pischke's (AP) partial F (Angrist and Pischke, (2009)), which are informative about the presence of weak instruments. Finally, for all over-identified models we calculate the Hausman test of equality of OLS and IV estimates.

We use the HJ over-identification and the CD statistics to focus our analysis on specifications with stronger and more exogenous instruments. In general, the Anderson–Rubin and Hausman tests strongly support our identification strategy. Based on the Anderson–Rubin test, we reject under-identification in all specifications for Guatemala, while for the Philippines we reject under-identification in 96% of the specifications. The Hausman test rejects equality of OLS and IV estimates in 99% of the specifications with total energy and 90% of the specifications with protein and non-protein separate in Guatemala and 87% and 98%, respectively, for the Philippines. Finally, we calculate the AP partial F statistic for the energy coefficient ($\lambda_{energy}^h$ and $\lambda_{energy}^w$) from equation (5) and the protein ($\lambda_{prot}^h$ and $\lambda_{prot}^w$) and non-protein coefficients ($\lambda_{non\_prot}^h$ and $\lambda_{non\_prot}^w$) from equation (6). These statistics are useful to make comparisons across equations and variables, but do not provide formal statistical support against weak instruments, since there are no critical values available for them. In general, the results suggest that the instruments are stronger for Guatemala: the AP partial F tends to be over 30 for the protein coefficients and over 7 for energy and non-protein coefficients. For the Philippines, the AP partial F for the total energy coefficient tends to be over 20. However, it is mostly below 5 for the protein and non-protein coefficients, which suggests that instruments are weaker in the more general specification for the Philippines.[18] Despite these differences in AP statistics, results are broadly similar across countries, which suggests that we are identifying structural relationships between nutrients and anthropometrics.

Since each production function is estimated multiple times, we explore distributions of estimated coefficients rather than a single or small set of "preferred" specifications, allowing us to draw more general conclusions. We do not choose or define a preferred specification because there are no obvious criteria for doing so and because of the concern that any potential preferred specification would not be robust to changes in the set of instruments. Although *a priori* the instruments we propose are plausibly exogenous and strong, we put relatively more confidence in those instrument sets that better satisfy over-identification and weak instrument tests.

The results of each type of specification are presented in Tables 3 to 6 and Figures 1 to 3. In Tables 3 and 5, and Figure 1, we present the estimated overall energy coefficients. In Tables 4 and 6 (Panels A and B), and Figure 2, we present the estimated protein coefficients, and in Tables 4 and 6 (Panels C and D), and Figure 3, the estimated non-protein coefficients. Each table presents the $25^{th}$, $50^{th}$ and $75^{th}$ percentiles of the estimated coefficient distributions and, in the final two columns, the percentages of the coefficient estimates that are significantly ($p<0.05$) positive or negative. For each Panel in each table, the first row reports distributions for all estimated specifications and, in subsequent rows, for specifications that are over-identified, and for those that have HJ p-values>0.05 and CD statistics>1, 3, or 7 (provided there are more than 10 such specifications in each case).[19] These sets of specifications focus on results for which relatively strong and exogenous instruments are available. Figures 1 to 3 present point estimates (and associated 95% confidence intervals) for all specifications that have HJ p-values>0.05 and CD>1 (corresponding to the third rows in Tables 3 to 6). The scale of the x-axis corresponds to the natural logarithm of CD statistics and the y-axis the coefficient values.[20]

---

[18] Results available on request.
[19] Restricting the sample to those with HJ p-values>0.10 generates similar results; see Data Appendix Section 5.

To facilitate interpretation of the coefficient magnitudes, we simulate changes in height and weight when energy intakes increase *ceteris paribus*. For this exercise, we use the most restrictive specifications with CD>7 (or CD>3 if there are fewer than ten specifications with CD>7) and HJ p-values>0.05. Within that set of specifications, we select the median coefficient and simulate effects of increasing energy intakes by 300 kcal per day, protein intakes by 10 grams per day, or non-protein intakes by 250 kcal per day. Each of these is approximately one SD of respective intakes of 18-month old infants in both countries. This hypothetical daily increase is then multiplied by 90 in Guatemala and by 60 in the Philippines to approximate total intakes for a given measurement period, and then multiplied by corresponding coefficients to obtain anthropometric changes. We call this exercise median prediction.

4.2. Guatemala

Table 3 summarizes for Guatemala distributions of coefficient estimates on total energy in the height and weight equations, and Figures 1A and 1B show the coefficients and confidence intervals for the corresponding specifications with CD>1. Total energy positively affects height and weight changes. These positive relationships are most evident for specifications with relatively stronger and more exogenous instruments. Our findings are consistent with previous literature that uses stronger identification assumptions estimating similar relationships from the same data sources (Habicht et al. (1995) and Griffen (2016)).

---

[20] The number of observations used varies for each specification. In the Data Appendix Section 4, we show that the results do not depend on the number of observations used.

For height in Guatemala, estimated coefficients on total energy are positive in the vast majority of cases, positive and significant ($p< 0.05$) in 35% of cases, and never negative and significant. The positive relationship is more robust when we consider specifications with relatively stronger and more exogenous instruments, according to the tests. Restricting to over-identified specifications in which HJ p-values>0.05 and CD>3, total energy coefficient estimates are positive and significant 57% of the time. To provide further interpretation of the magnitude of the coefficients, we calculate the median prediction (Section 4.1), taking the median coefficient of the specifications with CD>3; we calculate the effect of increasing energy per day by 300 kcal. For Guatemala, this implies a 0.62 cm predicted change in height.

For weight production functions, estimated coefficients on total energy are positive and significant for 36% of specifications, and are never significantly negative. Specifications with higher CD statistics have larger proportions of positive significant coefficient estimates. Figure 1B shows that while there are fewer specifications with higher CD statistic levels compared to the height model, for those with stronger instruments, the estimates are generally positive. The median prediction exercise indicates increasing energy intake by 300 kcal per day yields a predicted 620 grams change in weight.

Next, we consider the roles of protein and non-protein energy separately in the growth model. Proteins robustly and positively affect growth in height and weight in Guatemala, but the relationship of non-proteins (after controlling for protein) with these anthropometric measures is non-positive.

Panel A of Table 4 (and Figure 2A) shows that for 53% of all specifications, protein coefficient estimates are positive and significant. In specifications with CD>3, the estimates are always positive and significant. In specifications with stronger instruments, the estimated coefficient dispersion (i.e., the distance between the 25$^{th}$ and 75$^{th}$ percentiles) decreases; for specifications with CD>1 the ratio of the coefficients in the 75$^{th}$ and 25$^{th}$ percentiles is 1.3, while for the specifications with CD>3 the ratio is 1.06. Our median prediction exercise indicates that if protein were to increase by 10 grams per day, the predicted change in height is 0.39 cm.

For weight change (Panel B of Table 4 and Figure 2B), we find an even more robust pattern for proteins. In nearly all specifications (92%), protein coefficient estimates are positive and significant, and for specifications with CD>1, they are always positive and significant. For all specifications, the estimate at the 75$^{th}$ percentile is only 1.2 times larger than that at the 25$^{th}$ percentile. This pattern of stability and significance of coefficient estimates also can be seen in Figure 2B where the dispersion of the estimated coefficients is small, and there is a clear pattern of positive and significant effects of protein intake on weight growth. An increment in protein intake of 10 grams per day results in a predicted 195 gram change in weight.

By contrast, there is little evidence that energy from non-proteins affects changes in height and weight. Panel C in Table 4 and Figure 3A show that for Guatemala, in nearly all cases (98%) the estimated coefficient is insignificant in the height model. For the weight production function (Panel D of Table 4 and Figure 3B), the point estimates are never significant.

### 4.3. Philippines

Table 5 shows the distribution of the total energy coefficient estimates for the Philippines and Figures 1A and 1B the corresponding coefficients and confidence intervals for specifications with CD>1. As in Guatemala, positive relations are most evident for specifications with relatively stronger and more exogenous instruments. The positive impacts of total energy on height and weight are consistent with those found under somewhat stronger identification assumptions and using the same data, by Liu et al. (2009) and de Cao (2015).

Across all specifications summarized in the Panel A of Table 5, 13% have positive and significant coefficient estimates ($p<0.05$), while none have negative and statistically significant estimates. Restricting results to the 45 specifications with HJ test p-values>.05 and CD>7, 64% of estimated total energy coefficients are positive and significant. Specifications with higher CD statistics tend to have more concentrated coefficient estimate distributions. If daily energy intake increases by 300 kcal the predicted change in height is 0.18 cm.

For weight, evidence is similar regarding the role of total energy. The bottom panel of Table 5 indicates that for 15% of all the specifications in the Philippines, the estimated coefficient on total energy is positive and significant and never negative and significant. Specifications with the highest CD statistics tend to have larger shares of positive and significant coefficient estimates. Our median prediction results in a predicted change in weight of 37 grams.

Panel A of Table 6 (and Figure 2A) shows that for 39% of all specifications, protein coefficient estimates are positive and significant. While there are fewer specifications with strong instruments than in Guatemala, for specifications with CD>3, 100% of the coefficient estimates are positive and significant. In specifications with stronger instruments, the estimated coefficients dispersion decreases. Increasing protein consumption by 10 grams per day is predicted to 2.24 cm change in height.

For all specifications (Panel B of Table 6 and Figure 2B), 48% of estimated coefficients on protein for weight are positive and significant–100% in specifications with CD>3. Similar to Guatemala, coefficient estimate dispersion decreases with stronger instruments. Increasing protein consumption by 10 grams per day results in a predicted 703 grams change in weight.

Somewhat surprisingly, non-protein intakes are generally negatively related to both height and weight gain. For height, Panel C of Table 6 reports that 88% of the specifications with the strongest instruments (CD>3) yield negative and significant estimated coefficients. For weight, 100% of estimates in specifications with the strongest instruments are negative and significant.

These findings for non-protein energy for the Philippines are somewhat counter-intuitive, because they suggest that such energy intakes are detrimental to growth. Most individual foods (including those consumed in these regions during the study periods), however, include both proteins and non-proteins and virtually all diets do. Consequently, it is unlikely that actual intakes would change in a fashion that increased energy from non-proteins while simultaneously holding proteins constant. Since Filipino children's diets included both intakes, on net any negative effects of other macronutrient sources would have been partly or fully offset by protein effects. For example, not including breastmilk, at age 6 months, 93% of children had some protein consumption and from ages 14 to 24 months, all did. Moreover, at age 6 months 75% of children are breastfed, which also provides protein intakes. In section 4.5, we show that the model predicts that a dietary change (relatively rich in proteins but with some energy from other sources) indeed has positive effects on height and weight, despite negative coefficient estimates on non-proteins.

There are several potential explanations for the finding that non-proteins are less robustly related to anthropometrics than proteins. First, it is possible that energy from macronutrients other than proteins do not affect height and weight, at least aggregating the other macronutrients as we do. Second, it may be that non-linearities are not captured. For instance, it could happen that carbohydrates and fat need some proteins to have an effect on anthropometrics—if protein intakes are zero or very low, other intakes would not affect height and weight. Third, dietary changes after children stop breastfeeding can result in poorer quality diets, especially poor quality of carbohydrates and low micronutrient density, weakening any potential link to anthropometrics. Fourth, the available instruments simply may not be powerful enough to detect effects of other macronutrients; protein and non-protein intakes are highly correlated (even before instrumentation), making it difficult econometrically to identify their distinct effects; in that sense, Guatemala greatly benefits from the experimental Atole intervention, which provides a clear and strong exogenous variation for protein, though it is less powerful for other macronutrients.

### 4.4. Effects of other inputs and controls

In addition to the different nutrition intakes, our analysis provides estimates of the coefficients on lagged height, lagged weight, breastfeeding, and diarrhea. The results clearly indicate some catch-up height and weight growth. The lagged height coefficient is consistently negative and mostly significant in the change-in-height equation, indicating that shorter children at the end of one period tend to grow more in the next period. Similarly, the lagged weight coefficient is consistently negative and mostly significant in the weight equation so that lighter children at the end of one period gain more weight in the following period. With the caveat that the estimates for breastfeeding and diarrhea are potentially biased due to endogeneity, our coefficient estimates for number of days without diarrhea are consistently positive and significant for weight in both samples, suggesting that diarrhea has detrimental effects on weight gain as generally found in the literature. The coefficient estimates for breastfeeding are positive and mostly significant for Guatemala. In the Philippines, the coefficient estimates generally show a positive association between breastfeeding and height while the associations between breastfeeding and weight show no consistent pattern, similar to findings from Adair and Popkin (1996).[21]

### 4.5. Counterfactual Exercise: Increasing Nutritional Intakes

We next simulate the full effects of additional protein and non-protein intakes on child height and weight for the Philippines, complementing the simpler median predictions we used when interpreting individual coefficients. From the set of specifications with HJ p-values>0.05, we select the specification with the highest CD. The simulation is based on adding one egg per week to a child's diet, assuming no other changes in diet and no change in diarrhea. Eggs are good for such simulations. They were widely available in the localities where these studies are situated and are easily consumed by infants. They not only contain highly bioavailable protein, but also contain energy from other macronutrients, similar to many other naturally protein-rich foods. A medium (44 gram), whole raw egg contains on average 5.5 grams of protein and 40.9 calories from non-protein.[22, 23] Based on our parameter estimates, a child who consumed an additional egg per week on top of existing diet, for 18 months–from 6 to 24 months of age–would gain an additional 0.72 cm in height and 265 grams in weight.

---

[21] All results available on request.

## 5. Conclusions

Arimond and Ruel (2004) described associations between children's dietary diversity and their height. We build on their insights, examining effects of diet and particularly diet composition on height and weight growth for children between ages 6 and 24 months, giving special attention to differences between diets rich and poor in proteins. We improve upon previous literature by making weaker identifying assumptions, considering two important anthropometric measures—height and weight, investigating the robustness of our results to the use of a number of different instruments, and separately investigating the effects of energy from proteins and from non-proteins while controlling for breastfeeding and diarrhea. We take advantage of two rich databases, one for Guatemala and the other for the Philippines, which have longitudinal information on height, weight, and protein and energy intakes with high frequencies of observations. IV estimation strategies are used to overcome endogeneity and measurement error problems, using food prices and, in the case of Guatemala, a randomized nutritional intervention, as instruments. Because there are many instruments and instrument combinations available, we present results that comprehensively summarize these combinations rather than selecting only a single set of instruments. Our findings indicate that increasing energy intake increases both height and weight in both countries. But the source of that energy, protein versus non-protein, matters. In these poor populations characterized by high levels of chronic undernutrition, increases in protein intake drive increases in child height and weight.

---

[22] Agricultural Research Service of the United States Department of Agriculture. http://ndb.nal.usda.gov/ndb/foods/show/112 accessed on 17th September 2014.
[23] If households were to purchase the eggs, the cost would have been ~0.37% of the annual average income.

These results provide evidence on an important puzzle in the literature while pointing to possible modifications to interventions designed to improve children's nutritional status. A systematic review by Manley, Gitter and Slavchevska (2013) using meta-analysis techniques shows that while the average impact of income transfers from social protection programs on height-for-age is positive, effect sizes are small and not statistically significant. If households use these transfers largely to increase the quantity of calories consumed, if the increases in protein consumption is small in magnitude, or if these proteins are not allocated to children, then our results suggest that such transfers will have little impact on child height—precisely what Manley, Gitter and Slavchevska (2013) find. Headey and Hoddinott (2015) examine impacts of Green Revolution-induced increases in rice productivity on children's anthropometric status. They find no impact of these on child height, results also consistent with what we observe here. Our findings, in conjunction with these other studies, suggest that interventions designed to increase household incomes may only improve children's nutritional status when they are linked to mechanisms that also improve the quality of children's diets. Such interventions, e.g., linking nutritional behavior change communication to social protection interventions or "nutrition-sensitive agriculture" await further study.

**Funding**

The authors thank Grand Challenges Canada (Grant 0072-03), Bill & Melinda Gates Foundation (Global Health Grant OPP1032713), and the Eunice Shriver Kennedy National Institute of Child Health and Development (Grant R01 HD070993) for Financial Support. The funders have no involvement in the analysis and interpretation of the data, writing of the paper, or the decision to submit the paper for publication.

**Conflict of Interest**

There are no conflicts of interest.

**Acknowledgements**

This version of the paper has benefited with comments made by two anonymous referees and participants of the seminars at LACEA, PAA and University of Pennsylvania.

Table 1: Guatemala, Nutritional Outcomes and Inputs

|  | Height (cm) | Change in Height | Weight (grams) | Change in Weight | Total Energy (kcal) | Non-Protein (kcal) | Protein (grams) |
|---|---|---|---|---|---|---|---|
|  | mean (sd) | mean(sd) | mean(sd) | mean(sd) | mean(sd) | mean(sd) | mean(sd) |
| 6 months | 62.97 (2.38) | 5.19 (1.34) | 6871.99 (959.14) | 1424.26 (470.80) | 131.86 (149.30) | 113.82 (131.73) | 4.51 (5.45) |
| 9 months | 66.21 (2.69) | 3.46 (1.55) | 7516.29 (1085.65) | 698.85 (469.85) | 218.16 (193.81) | 191.06 (170.68) | 6.77 (6.85) |
| 12 months | 68.91 (3.00) | 2.96 (1.47) | 7979.84 (1147.19) | 500.85 (463.02) | 340.90 (232.77) | 301.12 (206.26) | 9.95 (7.78) |
| 15 months | 71.01 (3.21) | 2.40 (1.35) | 8292.93 (1117.15) | 461.96 (432.51) | 511.06 (245.47) | 451.65 (218.69) | 14.85 (8.39) |
| 18 months | 73.25 (3.36) | 2.29 (1.41) | 8712.95 (1118.61) | 431.83 (495.17) | 656.70 (271.83) | 581.27 (241.61) | 18.86 (9.29) |
| 21 months | 75.47 (3.47) | 2.33 (1.39) | 9186.83 (1129.93) | 505.42 (481.52) | 767.85 (293.42) | 678.13 (261.17) | 22.43 (10.08) |
| 24 months | 77.53 (3.55) | 2.23 (1.44) | 9752.69 (1168.04) | 604.67 (523.06) | 847.75 (303.51) | 747.65 (271.84) | 25.03 (10.37) |
| Observations | 3802 | 3802 | 3802 | 3802 | 3802 | 3802 | 3802 |

Continuation, Table 1: Philippines, Nutritional Outcomes and Inputs

|  | Height (cm) | Change in Height | Weight (grams) | Change in Weight | Total Energy (kcal) | Non-Protein (kcal) | Protein (grams) |
|---|---|---|---|---|---|---|---|
|  | mean(sd) | mean(sd) | mean(sd) | mean(sd) | mean(sd) | mean(sd) | mean(sd) |
| 6 months | 64.27 | 3.26 | 6856.72 | 736.13 | 204.93 | 182.57 | 5.59 |
|  | (2.57) | (1.66) | (903.05) | (415.00) | (249.68) | (221.73) | (7.52) |
| 8 months | 66.80 | 2.54 | 7302.63 | 440.11 | 285.67 | 254.65 | 7.76 |
|  | (2.71) | (1.42) | (964.47) | (383.29) | (279.54) | (246.70) | (8.99) |
| 10 months | 68.92 | 2.13 | 7642.79 | 338.95 | 349.93 | 312.18 | 9.44 |
|  | (2.80) | (1.39) | (1028.15) | (402.86) | (300.60) | (264.36) | (10.12) |
| 12 months | 70.72 | 1.82 | 7948.05 | 300.68 | 407.33 | 362.27 | 11.27 |
|  | (2.96) | (1.29) | (1079.27) | (391.39) | (310.79) | (273.21) | (10.56) |
| 14 months | 72.29 | 1.58 | 8225.85 | 278.16 | 477.29 | 423.20 | 13.52 |
|  | (3.07) | (1.22) | (1115.39) | (377.09) | (325.60) | (284.74) | (11.56) |
| 16 months | 73.73 | 1.45 | 8512.16 | 283.81 | 540.50 | 479.13 | 15.34 |
|  | (3.24) | (1.19) | (1111.83) | (389.74) | (328.22) | (285.88) | (12.22) |
| 18 months | 75.12 | 1.43 | 8797.30 | 286.79 | 589.04 | 521.87 | 16.79 |
|  | (3.38) | (1.23) | (1143.54) | (392.57) | (334.33) | (291.48) | (12.43) |
| 20 months | 76.50 | 1.42 | 9104.64 | 316.95 | 640.35 | 567.20 | 18.29 |
|  | (3.51) | (1.29) | (1177.77) | (397.24) | (347.83) | (303.38) | (12.74) |
| 22 months | 77.73 | 1.30 | 9436.95 | 338.55 | 681.66 | 603.35 | 19.58 |
|  | (3.61) | (1.33) | (1210.32) | (413.78) | (355.37) | (310.89) | (12.75) |
| 24 months | 79.13 | 1.43 | 9782.39 | 349.09 | 710.41 | 627.28 | 20.78 |
|  | (3.68) | (1.19) | (1233.11) | (418.55) | (354.38) | (309.68) | (12.99) |
| Observations | 24820 | 24820 | 24820 | 24820 | 24820 | 24820 | 24820 |

Table 2: Guatemala, Other Inputs

|  | Breastfed | Days with Diarrhea | Female | Time between measurement (days) | Age (days) |
|---|---|---|---|---|---|
|  | mean(sd) | mean(sd) | mean(sd) | mean(sd) | mean(sd) |
| 6 months | 0.99 | 6.52 | 0.51 | 91.95 | 182.62 |
|  | (0.12) | (12.44) | (0.50) | (5.09) | (3.66) |
| 9 months | 0.97 | 9.43 | 0.51 | 95.15 | 273.37 |
|  | (0.17) | (15.80) | (0.50) | (20.52) | (4.17) |
| 12 months | 0.92 | 12.18 | 0.50 | 96.82 | 364.59 |
|  | (0.28) | (16.08) | (0.50) | (24.77) | (4.88) |
| 15 months | 0.81 | 12.60 | 0.54 | 94.26 | 456.95 |
|  | (0.40) | (15.63) | (0.50) | (16.29) | (4.08) |
| 18 months | 0.59 | 11.43 | 0.53 | 94.48 | 547.98 |
|  | (0.49) | (15.77) | (0.50) | (18.90) | (3.51) |
| 21 months | 0.34 | 9.97 | 0.53 | 95.75 | 638.72 |
|  | (0.47) | (14.90) | (0.50) | (26.05) | (3.29) |
| 24 months | 0.18 | 7.57 | 0.53 | 100.14 | 730.64 |
|  | (0.38) | (13.65) | (0.50) | (34.32) | (3.21) |
| Observations | 3802 | 3802 | 3802 | 3802 | 3802 |

Continuation, Table 2: Philippines, Other Inputs

|  | Breastfed | Days with Diarrhea | Female | Time between measurement (days) | Age (days) |
|---|---|---|---|---|---|
|  | mean(sd) | mean(sd) | mean(sd) | mean(sd) | mean(sd) |
| 6 months | 0.76 | 1.54 | 0.53 | 61.77 | 186.41 |
|  | (0.43) | (2.76) | (0.50) | (8.92) | (6.03) |
| 8 months | 0.72 | 4.38 | 0.53 | 60.23 | 246.59 |
|  | (0.45) | (4.23) | (0.50) | (5.84) | (5.57) |
| 10 months | 0.68 | 3.11 | 0.53 | 62.03 | 307.98 |
|  | (0.47) | (2.35) | (0.50) | (8.61) | (6.03) |
| 12 months | 0.62 | 2.38 | 0.53 | 61.72 | 369.10 |
|  | (0.49) | (2.37) | (0.50) | (8.93) | (6.36) |
| 14 months | 0.53 | 5.25 | 0.53 | 61.48 | 430.07 |
|  | (0.50) | (5.15) | (0.50) | (8.55) | (6.46) |
| 16 months | 0.44 | 4.98 | 0.53 | 61.36 | 490.90 |
|  | (0.50) | (5.06) | (0.50) | (8.92) | (6.47) |
| 18 months | 0.34 | 1.99 | 0.53 | 61.54 | 551.72 |
|  | (0.47) | (2.35) | (0.50) | (9.56) | (6.16) |
| 20 months | 0.26 | 6.22 | 0.53 | 61.45 | 612.72 |
|  | (0.44) | (6.25) | (0.50) | (8.86) | (6.48) |
| 22 months | 0.19 | 2.78 | 0.53 | 60.83 | 673.14 |
|  | (0.39) | (3.51) | (0.50) | (8.55) | (6.11) |
| 24 months | 0.14 | 1.83 | 0.53 | 61.59 | 734.06 |
|  | (0.34) | (2.73) | (0.50) | (9.03) | (6.33) |
| Observations | 24820 | 24820 | 24820 | 24820 | 24820 |

Table 3: Impact of total energy intake on change in heights and weights, Guatemala

Panel A: Height

|  | Total Energy # of sp. | Distribution of total energy coefficient | | | sig>0 %-Sig | sig<0 %-Sig |
|---|---|---|---|---|---|---|
|  |  | p25 | p50 | p75 |  |  |
| All IV | 546 | -0.0182 | 0.0099 | 0.0288 | 35 | 0 |
| All Over-Identified IV | 525 | -0.0179 | 0.0107 | 0.0289 | 36 | 0 |
| CD>1 P-val HJ>5 | 137 | -0.0090 | 0.0034 | 0.0170 | 15 | 0 |
| CD>3 P-val HJ>5 | 21 | 0.0009 | 0.0231 | 0.0438 | 57 | 0 |

Panel B: Weight

| All IV | 546 | -0.0061 | 0.0059 | 0.0159 | 36 | 0 |
|---|---|---|---|---|---|---|
| All Over-Identified IV | 525 | -0.0036 | 0.0060 | 0.0159 | 38 | 0 |
| CD>1 P-val HJ>5 | 129 | -0.0024 | 0.0050 | 0.0233 | 32 | 0 |
| CD>3 P-val HJ>5 | 36 | 0.0142 | 0.0230 | 0.0239 | 83 | 0 |

1, CD = Robust Kleibergen-Paap F statistic, P-value J = p-value of Hansen J stat *100

2, 1st column: # of specifications that meet criteria; 2nd-4th col: percentile of distribution of estimated coefficients

3, 5th (6th) column: percent of estimated coefficients that are positive (negative) and significant at 5% significance level

4, 1st row: all specifications; 2nd row: all over-identified specifications for which # of IVs ># of endogenous variables. Other rows include all specifications satisfying the indicated criteria based on the CD and HJ tests.

5, All specifications include breastfeeding, diarrhea, sex, age, and age squared as covariates and a seasonal dummy for the Philippines, and lagged height and lagged weight, both of which are treated as endogenous.

6, Height coefficients are divided by 1000 for presentation purposes

Table 4: Impact of protein and non-protein energy on change in heights and weights, Guatemala.

Panel A: Height (Protein)

|  | Protein # of esp. | Distribution of protein coefficient | | | sig>0 %-Sig | sig<0 %-Sig |
|---|---|---|---|---|---|---|
|  |  | p25 | p50 | p75 |  |  |
| All IV | 525 | 0.0666 | 0.1047 | 0.1293 | 53 | 0 |
| All Over-Identified IV | 448 | 0.0774 | 0.1044 | 0.1268 | 58 | 0 |
| CD>1 P-val HJ>5 | 163 | 0.0931 | 0.1067 | 0.1232 | 77 | 0 |
| CD>3 P-val HJ>5 | 48 | 0.1043 | 0.1079 | 0.1106 | 100 | 0 |

Panel B: Weight (Protein)

| All IV | 525 | 0.0541 | 0.0588 | 0.0632 | 92 | 0 |
|---|---|---|---|---|---|---|
| All Over-Identified IV | 448 | 0.0543 | 0.0586 | 0.0627 | 97 | 0 |
| CD>1 P-val HJ>5 | 347 | 0.0540 | 0.0571 | 0.0614 | 100 | 0 |
| CD>3 P-val HJ>5 | 132 | 0.0534 | 0.0542 | 0.0567 | 100 | 0 |

Panel C: Height (Non-protein)

|  | Non-Protein # of esp. | Distribution of non-protein coefficient | | | sig>0 %-Sig | sig<0 %-Sig |
|---|---|---|---|---|---|---|
|  |  | p25 | p50 | p75 |  |  |
| All IV | 525 | -0.0170 | -0.0045 | 0.0039 | 0 | 2 |
| All Over-Identified IV | 448 | -0.0161 | -0.0042 | 0.0033 | 0 | 1 |
| CD>1 P-val HJ>5 | 163 | -0.0136 | -0.0053 | 0.0018 | 0 | 3 |
| CD>3 P-val HJ>5 | 48 | -0.0059 | -0.0028 | 0.0016 | 0 | 0 |

Panel D: Weight (Non-protein)

| All IV | 525 | -0.0019 | -0.0012 | -0.0005 | 0 | 0 |
|---|---|---|---|---|---|---|
| All Over-Identified IV | 448 | -0.0018 | -0.0012 | -0.0006 | 0 | 0 |
| CD>1 P-val HJ>5 | 347 | -0.0016 | -0.0012 | -0.0006 | 0 | 0 |
| CD>3 P-val HJ>5 | 132 | -0.0013 | -0.0011 | -0.0006 | 0 | 0 |

See Table 3 notes

Table 5: Impact of total energy intake on change in heights and weights, Philippines

Panel A: Height (See Figure 1A)

|  | Total Energy # of esp. | Distribution of total energy coefficient | | | sig>0 %-Sig | sig<0 %-Sig |
| --- | --- | --- | --- | --- | --- | --- |
|  |  | p25 | p50 | p75 |  |  |
| All IV | 602 | -0.0039 | 0.0069 | 0.0166 | 13 | 0 |
| All Over-Identified IV | 602 | -0.0039 | 0.0069 | 0.0166 | 13 | 0 |
| CD>1 P-val HJ>5 | 313 | -0.0174 | 0.0067 | 0.0147 | 18 | 0 |
| CD>3 P-val HJ>5 | 118 | 0.0035 | 0.0087 | 0.0140 | 37 | 0 |
| CD>7 P-val HJ>5 | 45 | 0.0076 | 0.0098 | 0.0123 | 64 | 0 |

Panel B: Weight (See Figure 1B)

|  | Total Energy # of esp. | Distribution of total energy coefficient | | | sig>0 %-Sig | sig<0 %-Sig |
| --- | --- | --- | --- | --- | --- | --- |
|  |  | p25 | p50 | p75 |  |  |
| All IV | 602 | 0.0013 | 0.0044 | 0.0220 | 15 | 0 |
| All Over-Identified IV | 602 | 0.0013 | 0.0044 | 0.0220 | 15 | 0 |
| CD>1 P-val HJ>5 | 284 | 0.0013 | 0.0058 | 0.0229 | 7 | 0 |
| CD>3 P-val HJ>5 | 65 | 0.0024 | 0.0063 | 0.0152 | 15 | 0 |
| CD>7 P-val HJ>5 | 15 | 0.0013 | 0.0020 | 0.0031 | 33 | 0 |

See Table 3 notes

Table 6: Impact of protein and non-protein energy on change in heights and weights, Philippines

Panel A: Height (See Figure 2A)

|  | Protein | Distribution of protein coefficient | | | sig>0 | sig<0 |
|---|---|---|---|---|---|---|
|  | # of esp. | p25 | p50 | p75 | %-Sig | %-Sig |
| All IV | 602 | 0.6826 | 1.0868 | 1.6848 | 39 | 0 |
| All Over-Identified IV | 448 | 0.7758 | 1.1194 | 1.7353 | 46 | 0 |
| CD>1 P-val HJ>5 | 248 | 0.8633 | 1.1247 | 1.4188 | 77 | 0 |
| CD>3 P-val HJ>5 | 16 | 0.6947 | 0.9324 | 1.0274 | 100 | 0 |

Panel B: Weight (See Figure 2B)

| All IV | 602 | 0.2972 | 0.3887 | 0.4818 | 48 | 0 |
|---|---|---|---|---|---|---|
| All Over-Identified IV | 448 | 0.3145 | 0.3991 | 0.4813 | 56 | 0 |
| CD>1 P-val HJ>5 | 242 | 0.3185 | 0.3766 | 0.4406 | 90 | 0 |
| CD>3 P-val HJ>5 | 16 | 0.2631 | 0.2929 | 0.3110 | 100 | 0 |

Panel C: Height (See Figure 3A)

|  | Non-Protein | Distribution of non-protein coefficient | | | sig>0 | sig<0 |
|---|---|---|---|---|---|---|
|  | # of esp. | p25 | p50 | p75 | %-Sig | %-Sig |
| All IV | 602 | -0.2792 | -0.1739 | -0.0943 | 0 | 32 |
| All Over-Identified IV | 448 | -0.2795 | -0.1789 | -0.1110 | 0 | 39 |
| CD>1 P-val HJ>5 | 248 | -0.2362 | -0.1777 | -0.1269 | 0 | 67 |
| CD>3 P-val HJ>5 | 16 | -0.1564 | -0.1283 | -0.0912 | 0 | 88 |

Panel D: Weight (See Figure 3B)

| All IV | 602 | -0.0754 | -0.0592 | -0.0433 | 0 | 38 |
|---|---|---|---|---|---|---|
| All Over-Identified IV | 448 | -0.0748 | -0.0606 | -0.0463 | 0 | 47 |
| CD>1 P-val HJ>5 | 242 | -0.0676 | -0.0577 | -0.0475 | 0 | 80 |
| CD>3 P-val HJ>5 | 16 | -0.0460 | -0.0433 | -0.0379 | 0 | 100 |

See Table 3 notes

Figure 1: Total Energy Coefficients

1A: Change in Height: Total Energy Coefficients

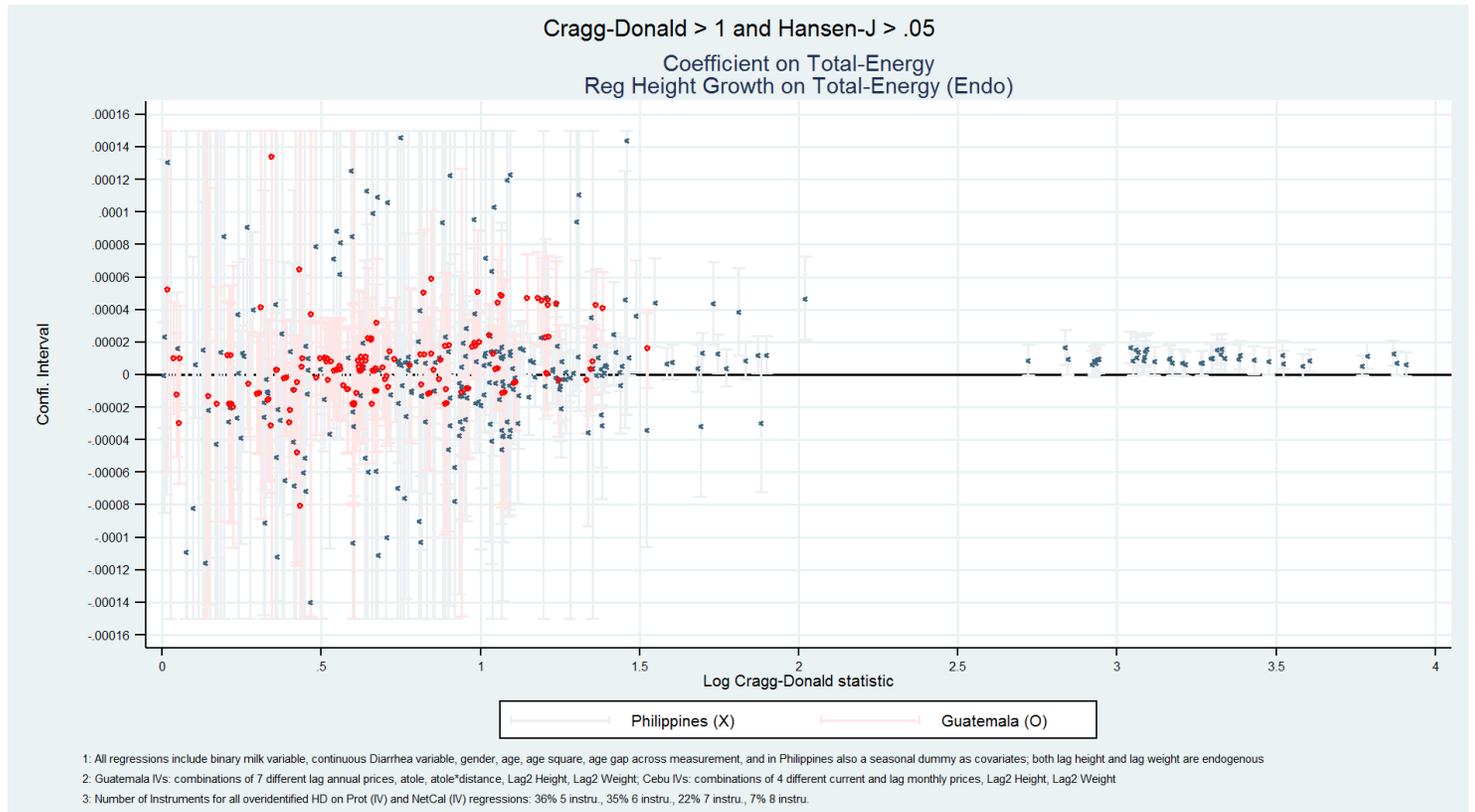

1B: Change in Weight: Total Energy Coefficients

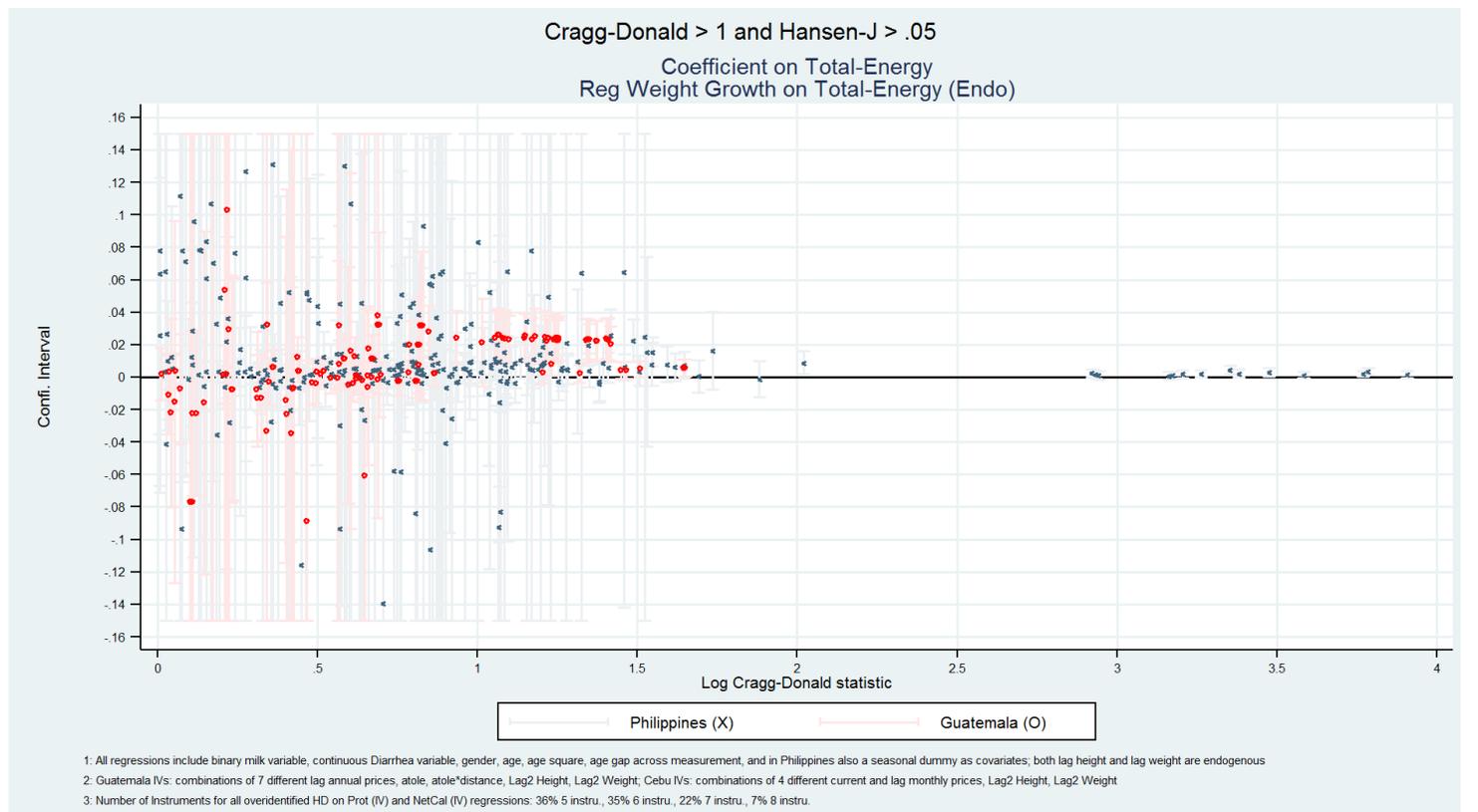

Figure 2: Protein Coefficients

2A: Change in Height: Protein Coefficients

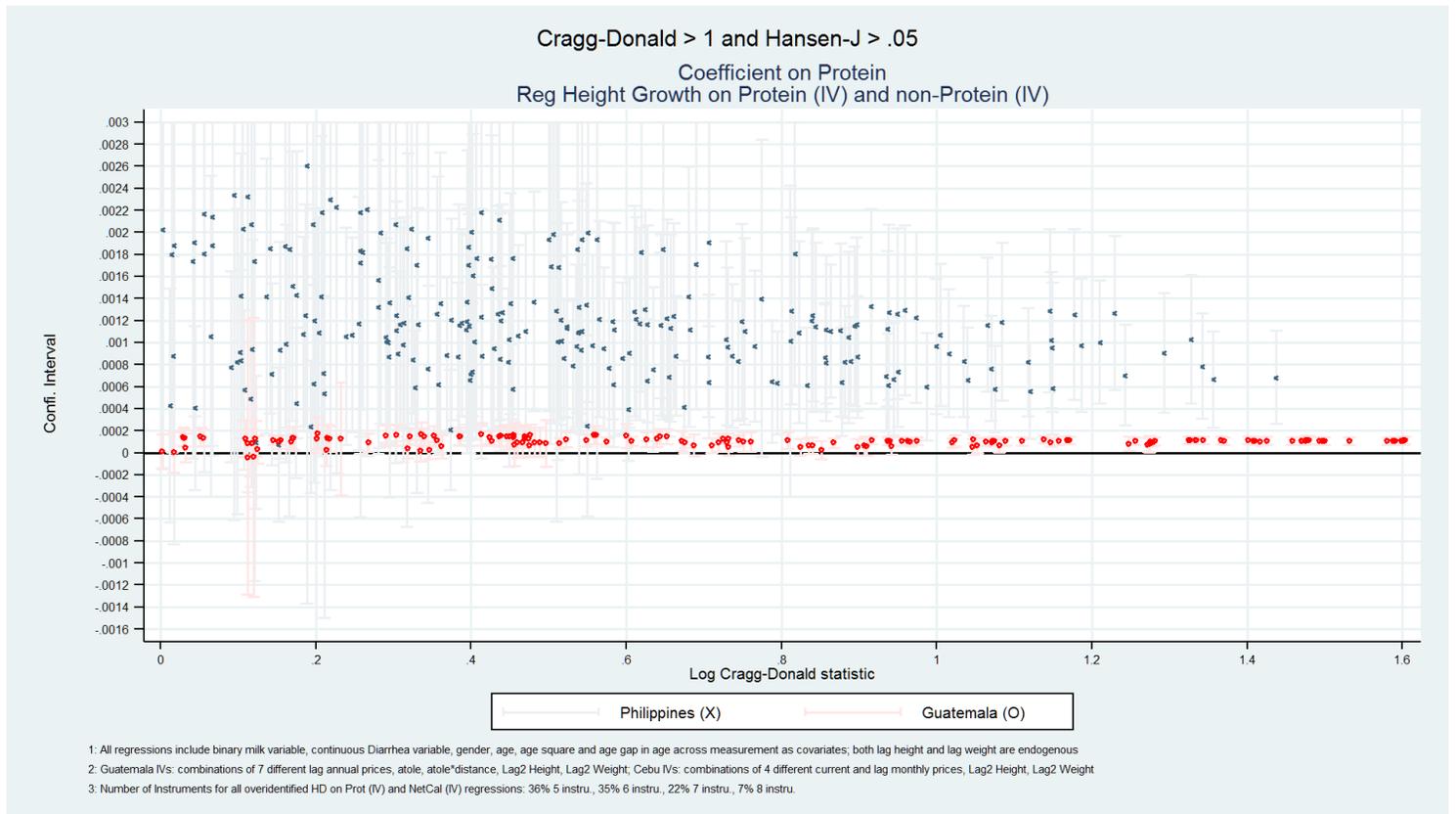

2B: Change in Weight: Protein Coefficients

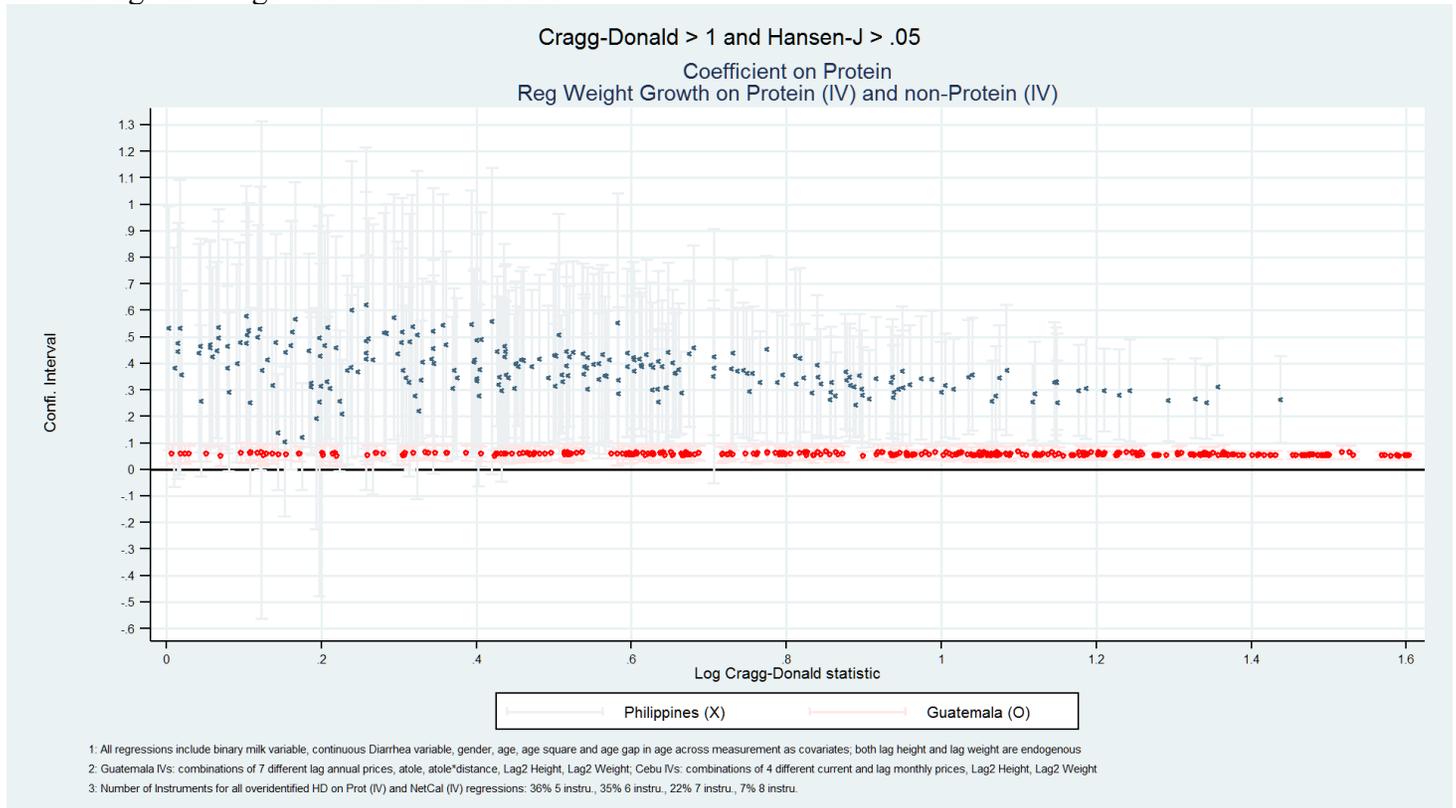

# Figure 3: Non-Protein Coefficients

## 3A: Change in Height: Non-Protein Coefficients

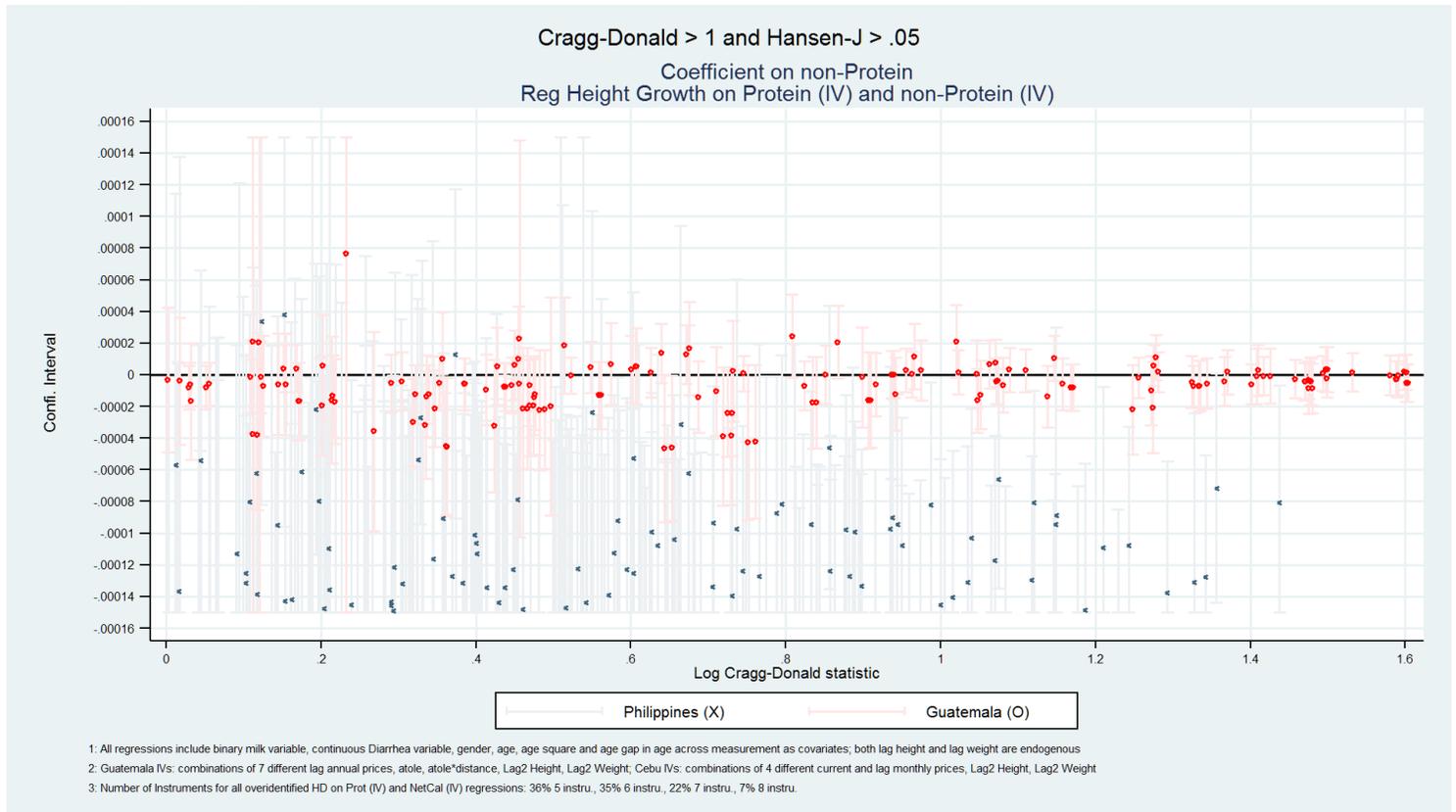

## 3B: Change in Weight: Non-Protein Coefficients

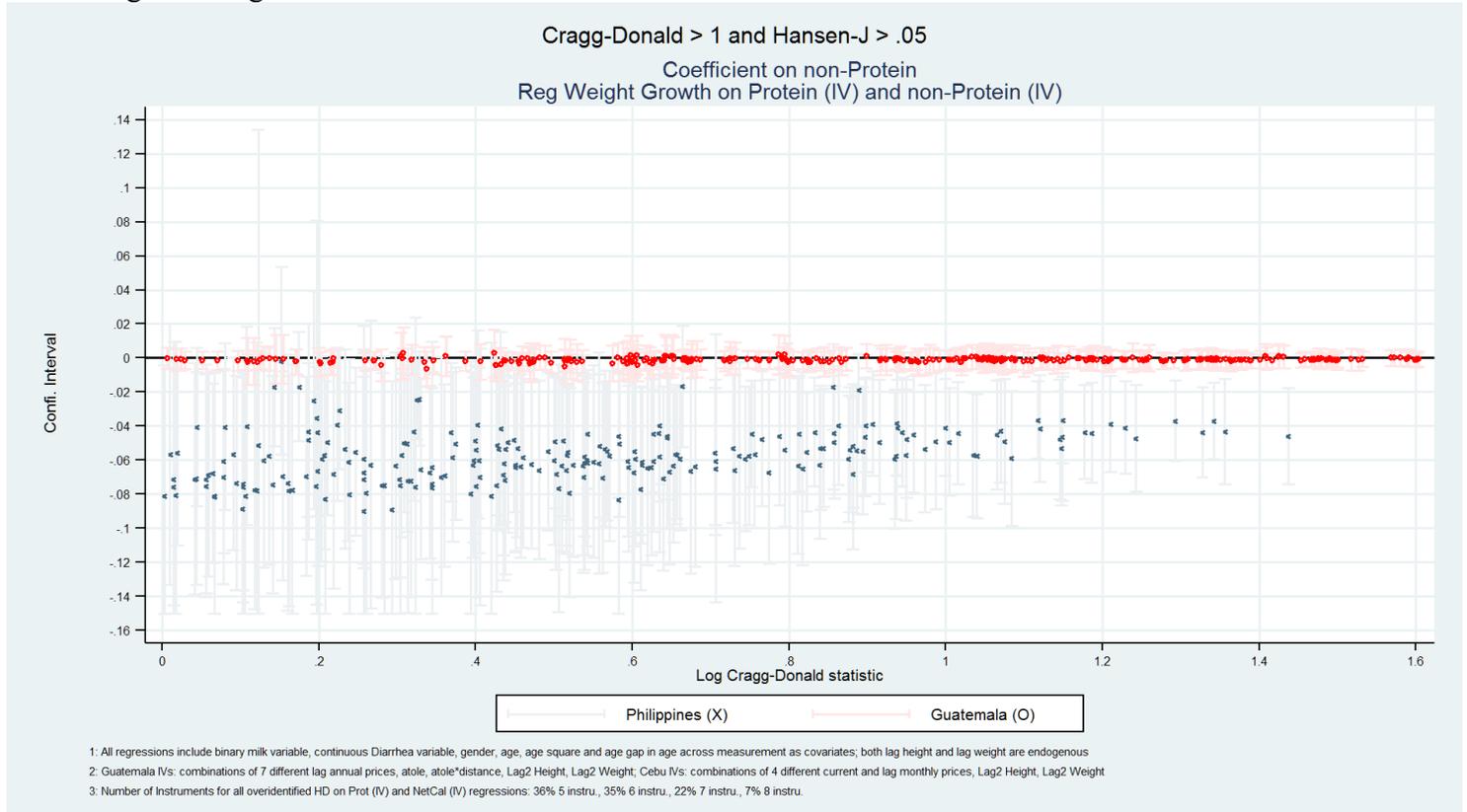

# DATA APPENDIX

## Early Life Height and Weight Production Functions with Endogenous Energy and Protein Inputs

**1. Imputation of missing protein and non-protein intakes in Guatemala**

The data for Guatemala included information on protein and non-protein intakes calculated from home consumption (based on the periodic home dietary 24- and 72-hour recall) and, separately, from supplement consumption (recorded daily at the feeding center), where the supplement was either Atole or Fresco. Supplements were prepared under careful oversight and intakes carefully measured (e.g., monitoring any amounts not consumed after a child was done) so their nutrient contents are well measured. We constructed total protein and non-protein intakes adding the intakes from both home and supplement consumption. In principle, intake data are available for children at ages: 3, 6, 9, 12, 15, 18, 21, 24, 30, 36, 42, 48, 60, 72 and 84 months. As described in the text, average daily intakes from home consumption were calculated for a given period (e.g., 3-6 months) by averaging the intakes at the two endpoints and then total intakes from home consumption were calculated by multiplying this daily average by the exact number of days between measurements. Supplement intakes were measured on a daily basis and were thus summed over the period. Last, intakes from both sources were totaled.

For some children at some ages, however, these total intakes are unavailable (either because an intake measure from home diet or from supplement was missing). We estimated flexible individual-level fixed-effects models for protein and non-protein intakes, separately, and used these models to impute some of the missing intakes. The model included as covariates age and year-of-birth dummies as shown in Table 1. Although our main analyses in the paper are from 6-24 months, the imputation model uses observation on all available ages (up to age 84 months) to improve the estimation of the individual effect.

Using the estimated individual fixed-effect models, we predicted intake values for a missing observation only when it was adjacent in time period to an observed intake, but we did not impute intakes if it was more than one measurement period away from an observed intake. For example, if there was missing information on intakes for a child at both ages 3, 6 and 9 months of age, but there was information at age 12, we predicted intakes at age 9, but not age 6, which remained as a missing observationa nd consequently the observation for that age for that child was not included in the estimation. In the previous example, if there was a measurement at age 3 months, we used the model to predict intake at age 6 months.

Appendix Table A1: Fixed-effect models for Protein and Non-Protein intakes in Guatemala

|  | (1) Protein | | (2) Non-Protein | |
| --- | --- | --- | --- | --- |
| Age in months | | | | |
| 3 months | omitted | | omitted | |
| 6 months | 12.24 | (4.938) | 76.11 | (4.261) |
| 9 months | 26.79 | (10.311) | 197.9 | (10.575) |
| 12 months | 41.81 | (15.041) | 324.6 | (16.208) |
| 15 months | 64.60 | (22.689) | 520.5 | (25.373) |
| 18 months | 81.40 | (26.904) | 625.8 | (28.712) |
| 21 months | 93.36 | (28.749) | 702.6 | (30.032) |
| 24 months | 107.2 | (30.375) | 787.8 | (30.971) |
| 30 months | 128.1 | (31.709) | 891.9 | (30.643) |
| 36 months | 142.3 | (30.461) | 984.4 | (29.254) |
| 42 months | 148.7 | (28.404) | 1009.9 | (26.781) |
| 48 months | 158.4 | (26.808) | 1092.9 | (25.674) |
| 60 months | 174.3 | (24.070) | 1179.4 | (22.614) |
| 72 months | 193.9 | (21.808) | 1287.1 | (20.092) |
| 84 months | 214.3 | (20.783) | 1432.2 | (19.277) |
| Birth year (=1) | | | | |
| 1968 | omitted | | omitted | |
| 1969 | -5.406 | (-1.063) | -63.16 | (-1.723) |
| 1970 | -12.96 | (-2.289) | -138.6 | (-3.397) |
| 1971 | -15.80 | (-2.433) | -189.8 | (-4.056) |
| 1972 | -32.21 | (-4.257) | -273.3 | (-5.013) |
| 1973 | -37.78 | (-4.324) | -276.6 | (-4.394) |
| 1974 | -45.70 | (-4.599) | -289.8 | (-4.049) |
| 1975 | -49.62 | (-4.436) | -285.8 | (-3.547) |
| 1976 | -48.10 | (-3.874) | -257.1 | (-2.874) |
| 1977 | -54.00 | (-4.027) | -274.6 | (-2.842) |
| Constant | 31.42 | (4.382) | 225.2 | (4.360) |
| Observations | 10117 | | 10117 | |

*t* statistics in parentheses

## 2. Diarrhea Models

The data from Guatemala includes detailed information on diarrhea including the number of days with diarrhea symptoms for each 15-day period during the first 24 months.[1] However, when one of these interviews was not conducted, the information on the number of days with diarrhea symptoms for that period are missing. To obtain an estimate of the total number of days with diarrhea for each growth period for the Guatemala analyses, we multiplied the average days of diarrhea of the observed period, with the length of the growth period, thereby assuming a similar incidence of diarrhea for the observed and unobserved days, as described in the text.

In the case of the Philippines, however, the data available for diarrhea only cover the seven days before each anthropometric measurement. To obtain a prediction of days with diarrhea for the entire 2-month growth period, we estimated count models using data from Guatemala, mimicking the information available in the Philippines, and then used those estimates to predict the number of days with diarrhea for each growth period in the Philippines. We took advantage detailed data from Guatemala with information on all the 15-days interviews to construct 2-month intervals for this estimation.

In specific, we estimated a count model to predict the total number of days with diarrhea for each two-month period; we used two-month periods to correspond directly to the data collection in the Philippines, which was every two months for each child. Since the prevalence of diarrhea varies with the age of the child, we estimated different models for each 2-month growth period, i.e., 0-2, 2-4, 4-6, etc. until 22-24 months of age.

There are a variety of different distributional functions that can be used for the count process when estimating a count model. With the aim of finding an accurate prediction model, we considered four different commonly used distributions: Poisson, Negative Binomial, Zero-Inflated Poisson and Zero-Inflated Negative Binomial model. All models were estimated using maximum likelihood.

---

[1] This information was organized by Humberto Méndez at INCAP during June-September 2013. The information that he used had 15-day intervals.

The Poisson distribution implies that the probability of observing $y$ days of diarrhea in a given two-month period is:

$$Pr[Y = y] = \frac{exp(\mu_i)\mu_i^y}{y!}$$

where $\mu_i = \exp(x_i'\beta)$ and $x_i$ are covariates. We used a limited set of covariates for each count model: the number of days with diarrhea in the last 15 days before a child is age $m$ months $(x_{m,15})$[2]; the average number of days of diarrhea of children of the same age in the same village over the same period; and indicators of the sex and birth order of the child. These covariates were chosen because of their availability across both samples, allowing us to apply the estimated coefficients from the model estimated with Guatemalan data to the Philippine data.

Since the objective of this estimation was to maximize the predictive power of the model, we evaluated different specifications of the covariates, always maintaining the basic variables described above. The additional covariates we included were variations of $x_{m,15}$; for instance, in the count model of the number of days with diarrhea during the growth period from 6 to 8 months of age, we can include not only the number of days with diarrhea 15 days before the child was age 8 months $x_{8,15}$, but also the number of days with diarrhea in the last 15 days previous to when the child was 6 months of age $x_{6,15}$ or the number of days with diarrhea in the last 15 days previous to measurement when the child was 10 months of age $x_{10,15}$ and so on.

To evaluate which model had the best fit we calculated the Bayesian information criteria (BIC) and an $R^2$ using the prediction of each model. To select the best specification and distribution we proceeded as follows. We compared (on the BIC and $R^2$) the specifications using the four potential distributions: Poisson, Negative Binomial count model, a Zero-Inflated Poisson model and a Zero-Inflated Negative

---

[2] This variable resembles the information available for the Philippines, which consists of the number of days with diarrhea 7 days before the child was age $m$ months.

Binomial model. At the same time, we calculated several specifications using different sets of the series of $(x_{m,15})$ from month 0 to month 24. Our results indicate first, that to predict the number of days with diarrhea in any growth period, all 15-day diarrhea measures (including both the "past" and the "future") should be included. Second, in terms of the BIC criteria, the Negative Binomial model outperformed the other distributions, but in terms of the $R^2$, the Poisson distribution was best. Then, to select the model that had the best fit for the Philippines between these two, we predicted the days without diarrhea using the Poisson and Negative binomial models, and correlated the predicted values with the actual information observed for the Philippines[3]. We found that the Poisson model had the highest correlation. Therefore we selected the Poisson distribution to predict the number of days without diarrhea for all two-month periods in the Philippines.

Following are the 12 models we used to predict days with diarrhea for the 12 growth periods in the Philippines.

---

[3] Remember that the data available in the Philippines allow constructing only the number of days with diarrhea in the last 7 days prior to the height and weight measurements, but we need to construct estimates for a two-month period.

Appendix Table A2: Estimation of Count Models of Number of Days with Diarrhea

|  | (1) |  |
|---|---|---|
|  | Number of Days with Diarrhea from age 0 mo to age 2 mo | |
| Days with diarrhea during 15 days previous to age: | | |
| 2 months | 0.287 | (29.761) |
| 4 months | 0.107 | (8.797) |
| 6 mo old | -0.0470 | (-3.629) |
| 8 mo old | -0.0241 | (-1.554) |
| 10 mo old | 0.0605 | (4.229) |
| 12 mo old | -0.0490 | (-3.921) |
| 14 mo old | -0.0623 | (-4.363) |
| 16 mo old | 0.0439 | (2.243) |
| 18 mo old | 0.0743 | (4.526) |
| 20 mo old | -0.0896 | (-4.178) |
| 22 mo old | 0.0343 | (1.903) |
| 24 mo old | -0.0695 | (-2.509) |
| Average days of diarrhea for children of same age, in same village, at same time | -0.558 | (-3.397) |
| Female (=1) | 0.123 | (0.976) |
| First born (=1) | -0.475 | (-2.590) |
| Second born (=1) | -2.391 | (-8.681) |
| Third born (=1) | 0.112 | (0.560) |
| Constant | 0.575 | (3.213) |
| Observations | 163 | |

$t$ statistics in parentheses

Count Model assuming Poisson Distribution. Maximum Likelihood Estimation.

Appendix Table A3: Number of Days with Diarrhea from age 2 mo to age 4 mo

|  | (1) |  |
|---|---|---|
|  | Number of Days with Diarrhea from age 2 mo to age 4 mo | |
| Days with diarrhea during 15 days previous to age: | | |
| 2 months | 0.116 | (18.447) |
| 4 months | 0.144 | (21.101) |
| 6 mo old | 0.0491 | (6.149) |
| 8 mo old | 0.0293 | (3.606) |
| 10 mo old | 0.0686 | (7.317) |
| 12 mo old | 0.00462 | (0.581) |
| 14 mo old | 0.0545 | (8.010) |
| 16 mo old | 0.00447 | (0.406) |
| 18 mo old | -0.0198 | (-2.004) |
| 20 mo old | -0.0507 | (-4.396) |
| 22 mo old | -0.108 | (-7.217) |
| 24 mo old | 0.0808 | (7.101) |
| Average days of diarrhea for children of same age, in same village, at same time | 0.677 | (6.731) |
| Female (=1) | 0.264 | (3.120) |
| First born (=1) | -0.795 | (-3.837) |
| Second born (=1) | 0.0881 | (0.882) |
| Third born (=1) | 0.375 | (2.477) |
| Constant | -0.736 | (-4.335) |
| Observations | 224 | |

*t* statistics in parentheses

Count Model assuming Poisson Distribution. Maximum Likelihood Estimation.

Appendix Table A4: Number of Days with Diarrhea from age 4 mo to age 6 mo

| | (1) |
|---|---|
| | Number of Days with Diarrhea from age 4 mo to age 6 mo |
| Days with diarrhea during 15 days previous to age: | |
| 2 months | 0.0293 (4.217) |
| 4 months | 0.0913 (14.936) |
| 6 mo old | 0.147 (24.269) |
| 8 mo old | -0.0139 (-1.806) |
| 10 mo old | 0.0494 (7.200) |
| 12 mo old | 0.0203 (3.001) |
| 14 mo old | 0.0612 (9.944) |
| 16 mo old | -0.00174 (-0.192) |
| 18 mo old | 0.0299 (5.003) |
| 20 mo old | 0.0680 (8.076) |
| 22 mo old | 0.0113 (1.393) |
| 24 mo old | -0.0320 (-2.896) |
| Average days of diarrhea for children of same age, in same village, at same time | 0.684 (8.752) |
| Female (=1) | 0.104 (1.512) |
| First born (=1) | -0.419 (-3.139) |
| Second born (=1) | 0.236 (2.619) |
| Third born (=1) | 0.153 (1.268) |
| Constant | -0.730 (-4.710) |
| Observations | 230 |

*t* statistics in parentheses

Count Model assuming Poisson Distribution. Maximum Likelihood Estimation.

Appendix Table A5: Number of Days with Diarrhea from age 6 mo to age 8 mo

|  | (1) |  |
|---|---|---|
|  | Number of Days with Diarrhea from age 6 mo to age 8 mo | |
| Days with diarrhea during 15 days previous to age: | | |
| 2 months | -0.00964 | (-1.283) |
| 4 months | -0.0136 | (-2.092) |
| 6 mo old | 0.0708 | (14.529) |
| 8 mo old | 0.123 | (21.737) |
| 10 mo old | -0.0151 | (-2.539) |
| 12 mo old | 0.0123 | (1.992) |
| 14 mo old | -0.0135 | (-2.045) |
| 16 mo old | 0.0464 | (5.867) |
| 18 mo old | 0.00665 | (1.193) |
| 20 mo old | 0.0281 | (3.608) |
| 22 mo old | 0.0417 | (5.826) |
| 24 mo old | 0.0337 | (3.732) |
| Average days of diarrhea for children of same age, in same village, at same time | 0.229 | (2.763) |
| Female (=1) | -0.294 | (-4.883) |
| First born (=1) | 0.296 | (3.002) |
| Second born (=1) | -0.138 | (-1.779) |
| Third born (=1) | 0.165 | (1.756) |
| Constant | 0.853 | (5.756) |
| Observations | 222 | |

*t* statistics in parentheses

Count Model assuming Poisson Distribution. Maximum Likelihood Estimation.

Appendix Table A6: Number of Days with Diarrhea from age 8 mo to age 10 mo

|  | (1) Number of Days with Diarrhea from age 8 mo to age 10 mo | |
|---|---|---|
| Days with diarrhea during 15 days previous to age: | | |
| 2 months | -0.00167 | (-0.255) |
| 4 months | 0.00166 | (0.231) |
| 6 mo old | 0.0137 | (2.459) |
| 8 mo old | 0.0289 | (5.435) |
| 10 mo old | 0.111 | (25.650) |
| 12 mo old | 0.0000986 | (0.019) |
| 14 mo old | 0.0444 | (8.275) |
| 16 mo old | 0.0232 | (3.595) |
| 18 mo old | 0.0158 | (2.983) |
| 20 mo old | -0.00989 | (-1.270) |
| 22 mo old | 0.0345 | (5.328) |
| 24 mo old | -0.000590 | (-0.069) |
| Average days of diarrhea for children of same age, in same village, at same time | 0.410 | (6.321) |
| Female (=1) | -0.196 | (-4.140) |
| First born (=1) | 0.134 | (1.642) |
| Second born (=1) | -0.139 | (-1.839) |
| Third born (=1) | 0.0317 | (0.375) |
| Constant | 0.600 | (3.727) |
| Observations | 222 | |

*t* statistics in parentheses

Count Model assuming Poisson Distribution. Maximum Likelihood Estimation.

Appendix Table A7: Number of Days with Diarrhea from age 10 mo to age 12 mo

|  | (1) |  |
|---|---|---|
|  | Number of Days with Diarrhea from age 10 mo to age 12 mo | |
| Days with diarrhea during 15 days previous to age: | | |
| 2 months | 0.0275 | (5.138) |
| 4 months | -0.00910 | (-1.216) |
| 6 mo old | -0.0183 | (-2.968) |
| 8 mo old | 0.0121 | (2.189) |
| 10 mo old | 0.0508 | (10.795) |
| 12 mo old | 0.0926 | (20.325) |
| 14 mo old | 0.0332 | (6.712) |
| 16 mo old | 0.0151 | (2.168) |
| 18 mo old | 0.00732 | (1.286) |
| 20 mo old | -0.0168 | (-2.087) |
| 22 mo old | 0.0599 | (8.622) |
| 24 mo old | 0.0151 | (1.776) |
| Average days of diarrhea for children of same age, in same village, at same time | 0.487 | (8.003) |
| Female (=1) | -0.00107 | (-0.023) |
| First born (=1) | -0.0393 | (-0.443) |
| Second born (=1) | 0.734 | (11.764) |
| Third born (=1) | -0.730 | (-5.507) |
| Constant | 0.392 | (2.838) |
| Observations | 232 | |

$t$ statistics in parentheses

Count Model assuming Poisson Distribution. Maximum Likelihood Estimation.

Appendix Table A8: Number of Days with Diarrhea from age 12 mo to age 14 mo

|  | (1) |  |
|---|---|---|
|  | Number of Days with Diarrhea from age 12 mo to age 14 mo | |
| Days with diarrhea during 15 days previous to age: | | |
| 2 months | 0.0448 | (8.110) |
| 4 months | -0.0419 | (-4.520) |
| 6 mo old | 0.0247 | (4.084) |
| 8 mo old | -0.0192 | (-2.854) |
| 10 mo old | 0.0196 | (3.386) |
| 12 mo old | 0.0382 | (7.374) |
| 14 mo old | 0.151 | (28.242) |
| 16 mo old | 0.00744 | (0.840) |
| 18 mo old | 0.0130 | (2.105) |
| 20 mo old | -0.0181 | (-1.839) |
| 22 mo old | 0.0284 | (3.824) |
| 24 mo old | 0.0517 | (5.846) |
| Average days of diarrhea for children of same age, in same village, at same time | 0.0626 | (0.922) |
| Female (=1) | 0.121 | (2.288) |
| First born (=1) | 0.316 | (3.767) |
| Second born (=1) | -0.124 | (-1.455) |
| Third born (=1) | 0.0627 | (0.569) |
| Constant | 0.982 | (6.751) |
| Observations | 225 | |

*t* statistics in parentheses

Count Model assuming Poisson Distribution. Maximum Likelihood Estimation.

Appendix Table A9: Number of Days with Diarrhea from age 14 mo to age 16 mo

(1)

Number of Days with Diarrhea from age 14 mo to age 16 mo

| Days with diarrhea during 15 days previous to age: | | |
|---|---:|---:|
| 2 months | 0.0271 | (4.449) |
| 4 months | -0.0429 | (-5.214) |
| 6 mo old | 0.0357 | (6.074) |
| 8 mo old | 0.0566 | (9.995) |
| 10 mo old | -0.0161 | (-2.649) |
| 12 mo old | -0.00774 | (-1.258) |
| 14 mo old | 0.0569 | (9.994) |
| 16 mo old | 0.133 | (21.063) |
| 18 mo old | 0.0108 | (1.904) |
| 20 mo old | 0.0281 | (4.062) |
| 22 mo old | 0.0555 | (8.296) |
| 24 mo old | -0.00892 | (-1.032) |
| Average days of diarrhea for children of same age, in same village, at same time | -0.0138 | (-0.210) |
| Female (=1) | -0.0117 | (-0.208) |
| First born (=1) | -0.468 | (-4.175) |
| Second born (=1) | -0.233 | (-2.903) |
| Third born (=1) | -0.0504 | (-0.495) |
| Constant | 1.201 | (9.461) |
| Observations | 231 | |

$t$ statistics in parentheses

Count Model assuming Poisson Distribution. Maximum Likelihood Estimation.

Appendix Table A10: Number of Days with Diarrhea from age 16 mo to age 18 mo

|  | (1) |  |
|---|---|---|
|  | Number of Days with Diarrhea from age 16 mo to age 18 mo | |
| Days with diarrhea during 15 days previous to age: | | |
| 2 months | 0.00473 | (0.595) |
| 4 months | -0.0179 | (-1.853) |
| 6 mo old | 0.00613 | (0.984) |
| 8 mo old | 0.0733 | (10.689) |
| 10 mo old | -0.0577 | (-8.005) |
| 12 mo old | 0.0109 | (1.704) |
| 14 mo old | 0.0143 | (1.972) |
| 16 mo old | 0.106 | (15.217) |
| 18 mo old | 0.116 | (19.520) |
| 20 mo old | -0.00579 | (-0.656) |
| 22 mo old | 0.0252 | (2.870) |
| 24 mo old | 0.0325 | (3.633) |
| Average days of diarrhea for children of same age, in same village, at same time | 0.504 | (6.553) |
| Female (=1) | 0.152 | (2.567) |
| First born (=1) | -0.129 | (-1.383) |
| Second born (=1) | -0.102 | (-1.265) |
| Third born (=1) | -0.189 | (-1.640) |
| Constant | 0.0817 | (0.478) |
| Observations | 228 | |

*t* statistics in parentheses

Count Model assuming Poisson Distribution. Maximum Likelihood Estimation.

Appendix Table A11: Number of Days with Diarrhea from age 18 mo to age 20 mo

|  | (1) |  |
|---|---|---|
|  | Number of Days with Diarrhea from age 18 mo to age 20 mo | |
| Days with diarrhea during 15 days previous to age: | | |
| 2 months | -0.0461 | (-5.830) |
| 4 months | 0.00105 | (0.126) |
| 6 mo old | 0.00917 | (1.306) |
| 8 mo old | 0.0310 | (4.103) |
| 10 mo old | 0.00640 | (1.007) |
| 12 mo old | -0.0314 | (-4.250) |
| 14 mo old | 0.0814 | (13.091) |
| 16 mo old | -0.0250 | (-3.031) |
| 18 mo old | 0.0907 | (16.077) |
| 20 mo old | 0.159 | (25.675) |
| 22 mo old | 0.0171 | (2.203) |
| 24 mo old | 0.0307 | (3.986) |
| Average days of diarrhea for children of same age, in same village, at same time | -0.0692 | (-0.722) |
| Female (=1) | -0.0129 | (-0.226) |
| First born (=1) | 0.350 | (3.806) |
| Second born (=1) | 0.0976 | (1.110) |
| Third born (=1) | 0.208 | (2.267) |
| Constant | 1.025 | (7.206) |
| Observations | 231 | |

*t* statistics in parentheses

Count Model assuming Poisson Distribution. Maximum Likelihood Estimation.

Appendix Table A12  Number of Days with Diarrhea from age 20 mo to age 22 mo

|  | (1) |  |
|---|---|---|
|  | Number of Days with Diarrhea from age 20 mo to age 22 mo | |
| Days with diarrhea during 15 days previous to age: |  |  |
| 2 months | 0.0186 | (2.305) |
| 4 months | -0.0185 | (-2.029) |
| 6 mo old | 0.0374 | (5.732) |
| 8 mo old | 0.0231 | (3.154) |
| 10 mo old | 0.0102 | (1.557) |
| 12 mo old | 0.0284 | (3.896) |
| 14 mo old | 0.0290 | (4.083) |
| 16 mo old | 0.0130 | (1.510) |
| 18 mo old | 0.0119 | (1.834) |
| 20 mo old | 0.0943 | (13.018) |
| 22 mo old | 0.150 | (22.769) |
| 24 mo old | 0.0342 | (4.147) |
| Average days of diarrhea for children of same age, in same village, at same time | 0.271 | (3.697) |
| Female (=1) | 0.0641 | (1.008) |
| First born (=1) | -0.417 | (-2.958) |
| Second born (=1) | 0.516 | (5.429) |
| Third born (=1) | -0.234 | (-2.297) |
| Constant | 0.334 | (2.956) |
| Observations | 223 |  |

*t* statistics in parentheses

Count Model assuming Poisson Distribution. Maximum Likelihood Estimation.

Appendix Table A13: Number of Days with Diarrhea from age 22 mo to age 24 mo

|  | (1) |  |
|---|---|---|
|  | Number of Days with Diarrhea from age 22 mo to age 24 mo | |
| Days with diarrhea during 15 days previous to age: | | |
| 2 months | 0.0363 | (3.757) |
| 4 months | -0.0254 | (-2.108) |
| 6 mo old | -0.0176 | (-1.920) |
| 8 mo old | 0.0564 | (6.174) |
| 10 mo old | -0.0124 | (-1.453) |
| 12 mo old | -0.0105 | (-1.093) |
| 14 mo old | -0.00228 | (-0.288) |
| 16 mo old | -0.0120 | (-1.159) |
| 18 mo old | 0.0441 | (5.108) |
| 20 mo old | 0.0167 | (2.038) |
| 22 mo old | 0.120 | (13.621) |
| 24 mo old | 0.187 | (26.108) |
| Average days of diarrhea for children of same age, in same village, at same time | 0.0989 | (1.057) |
| Female (=1) | 0.138 | (1.798) |
| First born (=1) | -0.399 | (-2.474) |
| Second born (=1) | -0.418 | (-3.476) |
| Third born (=1) | -0.861 | (-5.692) |
| Constant | 0.524 | (4.370) |
| Observations | 232 | |

*t* statistics in parentheses

Count Model assuming Poisson Distribution. Maximum Likelihood Estimation.

### 3. Food prices used in the instrument sets

*Guatemala*

For Guatemala, we use lagged annual prices of eggs, chicken, pork, beef, dry beans, corn, and rice, obtained from the FAO through its website FAOSTAT. All prices are available over the eight-year study period (and thus range from 1968 to 1976).

*Philippines*

For the Philippines, we use community-specific[4] prices collected as part of the broader study. Beginning in January 1983 (and ending in May 1986), enumerators visited two stores in each community, every other month, and collected prices (and quantity units) for a list of items. Not all items, however, were sold at each store at each visit. Consequently, there is not a complete set of prices for each item from each store (or even from each community in instances where no price was available from either store) in each measurement period.

1. Conversion to unit prices
    - For the same product, across different stores or visits, not all prices were collected in the same units (possible units included for example piece, hundred, pack, kg and chupa). Prices per unit also differed for goods for sale in different quantities.
    - All prices were converted to the value in Philippine Centavos for 100 grams of product.
2. Averaging, Outliers and Interpolation
    - When a unit price was available for both stores in a community at a given time, then the average of those prices was used for that community and time period. If only one unit price was available, then that price was used for that community and time period.
    - There were some examples of extreme values of unit prices, possibly due to incorrect unit conversions. A small percentage of these extreme values were dropped.
    - As the price surveys were administered during odd numbered months

---

[4] Communities correspond to what is known locally as the barangay—the smallest administrative division in the Philippines.

(January, March, etc.), prices for even numbered months were calculated as the average of two adjacent odd numbered months.

3. Our analyses use as instruments prices for dried fish, eggs, corn, and tomatoes. Although prices for many other food items also were collected (including chicken, beef, pork, fried fish, and cabbage), they were often missing. After interpolating for even months, the number of community-month observations on different prices are as follows, out of a maximum possible of 1353:

   Food items used as instruments
   - Corn (N= 835)
   - Dried Fish (N=760)
   - Eggs (N=783)
   - Tomatoes (N=794)

   Food items not used as instruments
   - C4 rice (N=497)
   - Wagwag rice (N=318)
   - Pork (N=262)
   - Beef (N=218)
   - Chicken (N=176)

All prices were inflated to year 2000 Philippine Centavos using monthly inflation rates. These were obtained from the website of Bangko Sentralng Pilipinas, the Central Bank of the Philippines: http://www.bsp.gov.ph/dbank_reports/Prices_1_rpt.asp?frequency=Monthly&range_from=1980&range_to=2011&conversion=None accessed on September 17th, 2014.

## 4. Specifications and number of observations

In this section, we examine whether the point estimates for the Philippines vary with the number of observations used; the driving factor behind the differing numbers of observations across specifications is the availability of food prices for the each particular set of instruments used, as described in the previous section. The number of observations ranges from approximately 7,000 to 15,000.

We present four figures. Figure 1 shows the protein coefficients and figure 2 the non-protein coefficients for the height production function. Figures 3 and 4 show the protein and non-protein coefficients, respectively, for the weight production function. We graph the point estimates (y-axis) against the number of observations (x-axis) for all specifications, dividing each set of results into two sub-panels. The left-side panel displays specifications with a CD < 3 and the right-side panel, with CD >3. The figures also indicate which specifications meet the 0.1 threshold for the p-value of the HJ as defined in the legend.

The graphs indicate that there does not appear to be any relationship between the point estimates and the number of observations. Nor does there appear to be evidence that they systematically vary according by the cut-off of 0.1 for the p-value of the HJ statistics.

Appendix Figure A1: Protein Coefficient in the Height Specification and Number of Observations

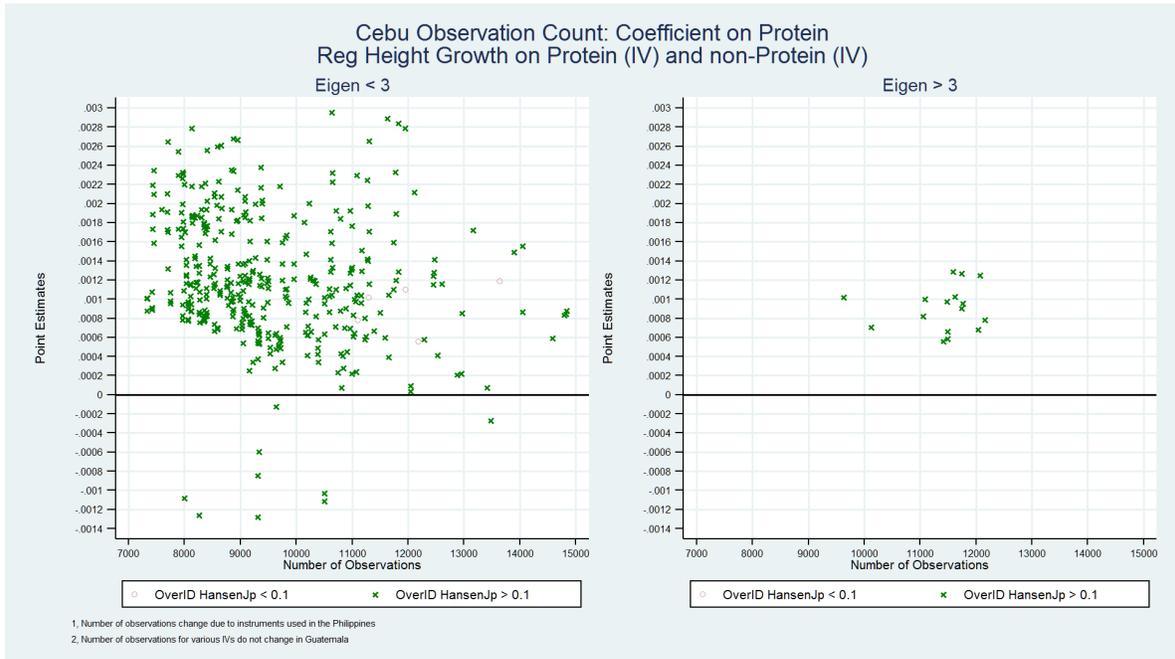

Appendix Figure A2: Non-Protein Coefficient in the Height Specification and Number of Observations

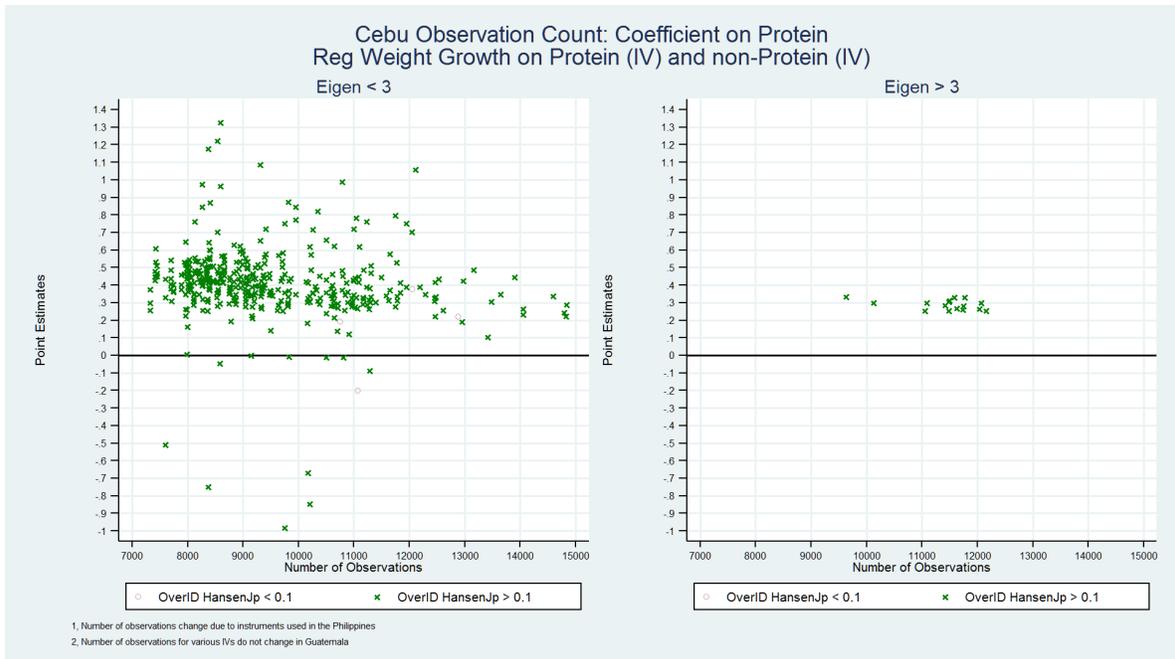

Appendix Figure A3: Protein Coefficient in the Weight Specification and Number of Observations

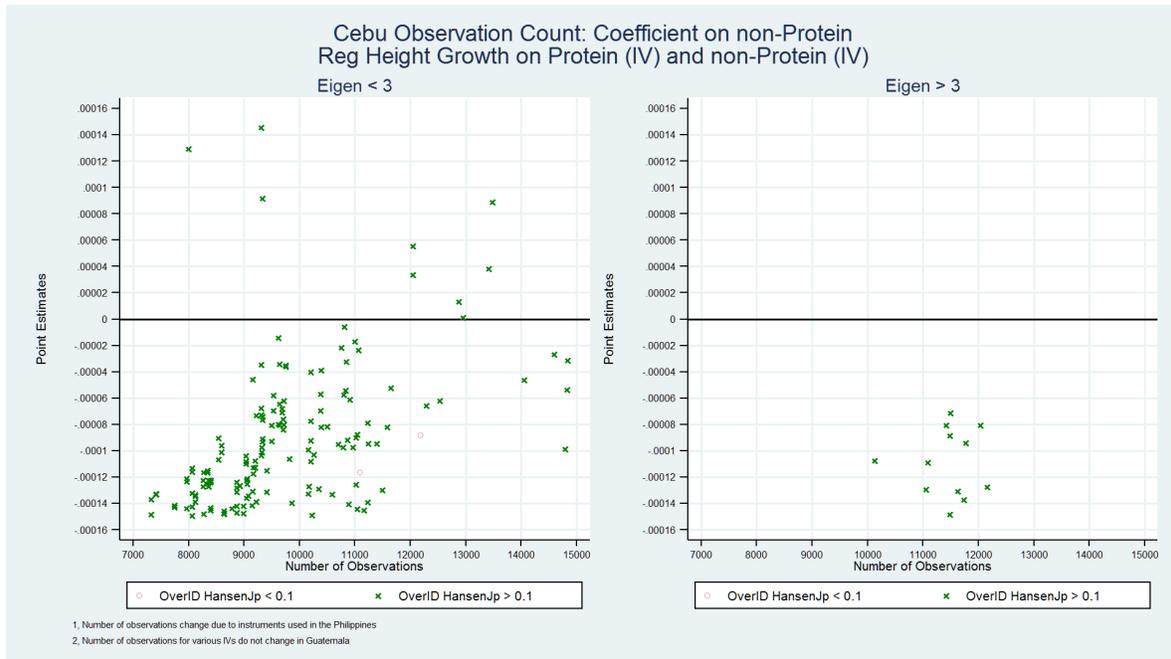

Appendix Figure A4: Non-Protein Coefficient in the Weight Specification and Number of Observations

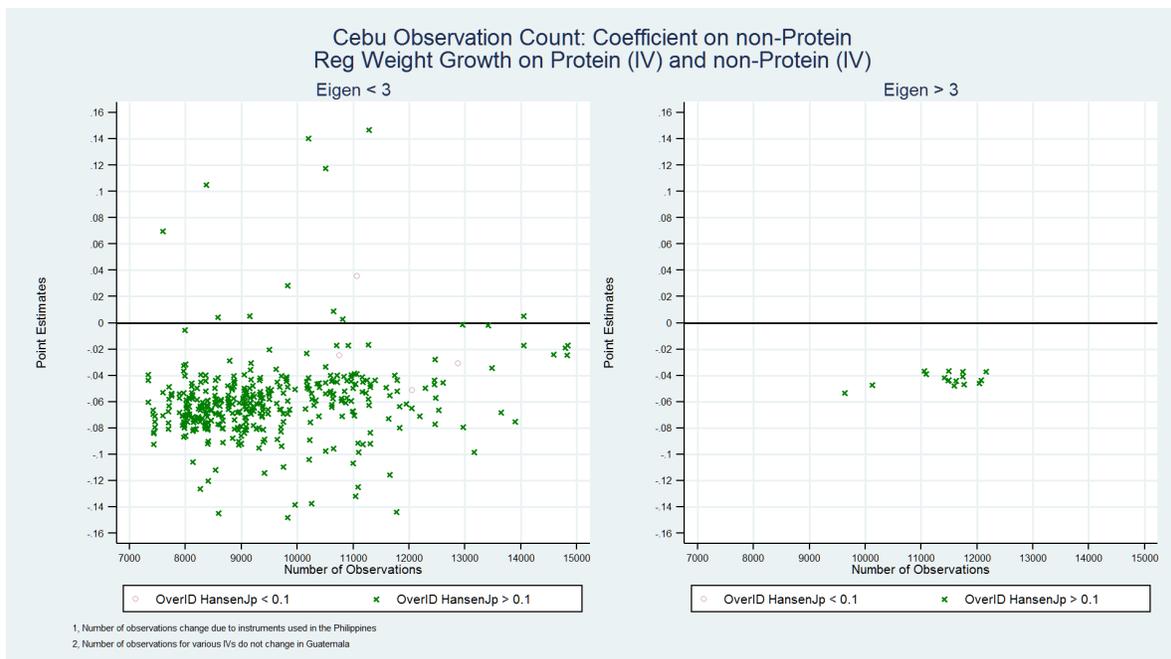

## 5. Height and Weight production functions

In this section, we present the set of Figures paralleling Figures 1 to 3 in the paper, but using a cutoff of 0.10 for the p-value of the Hansen-J test. Results are little changed from those with a cutoff of 0.05 discussed in the paper.

Figure A5: Total Energy Coefficient

Change in Height: Estimates $\lambda^h_{energy}$ from Equation (5) in the manuscript

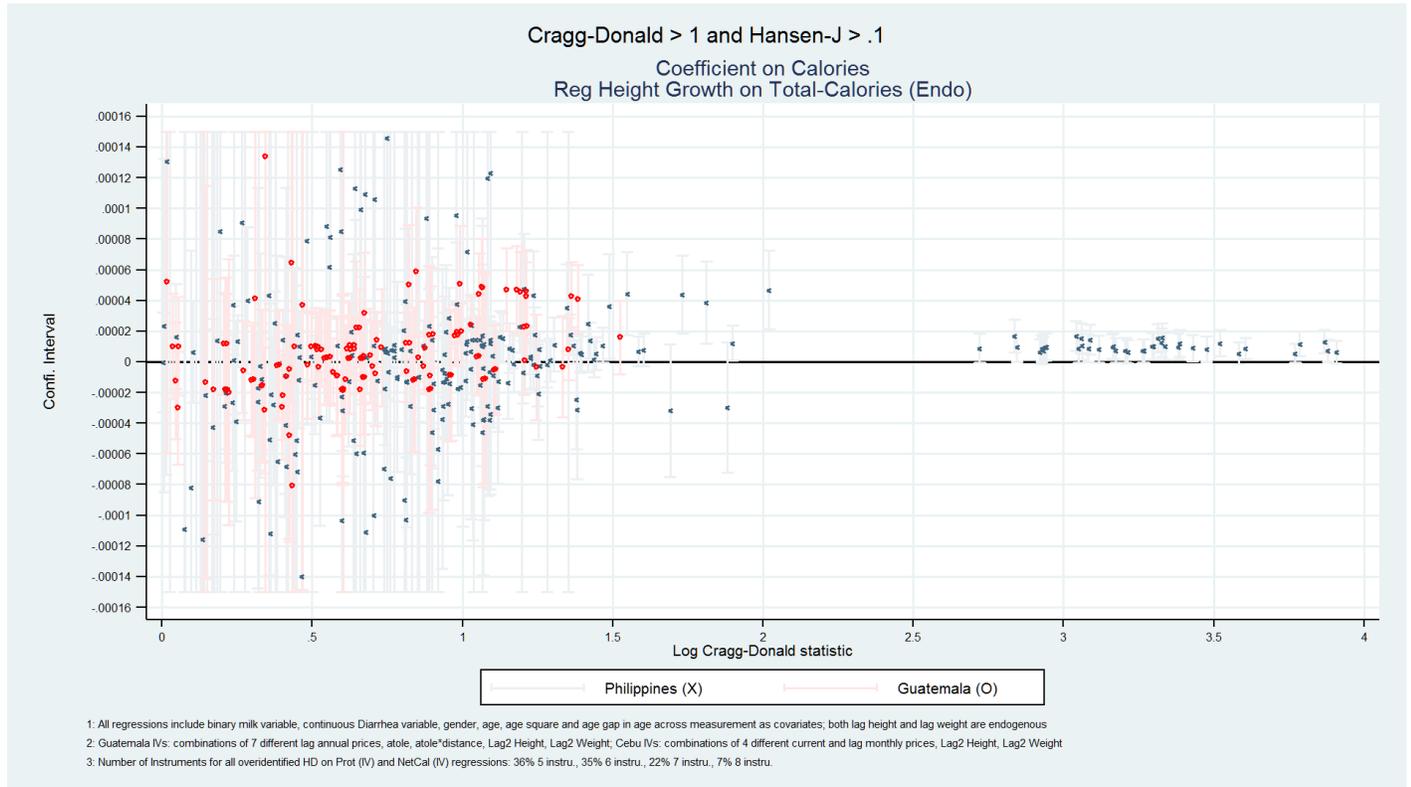

Change in Weight: Estimates $\lambda^w_{energy}$ Equation (5) in the manuscript

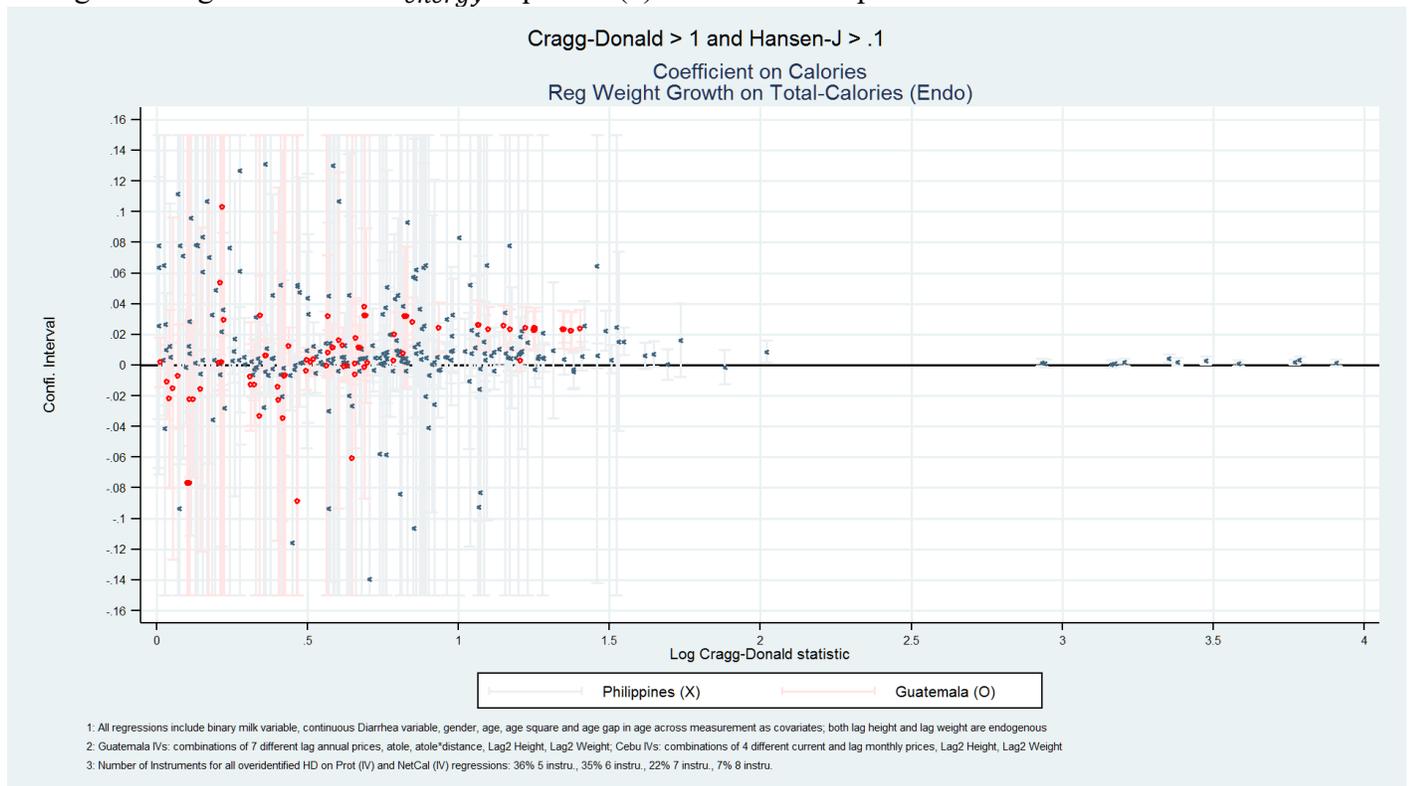

Figure A6: Protein Coefficient
Change in Height: Estimates $\lambda^h_{prot}$ from Equation (6) in the manuscript

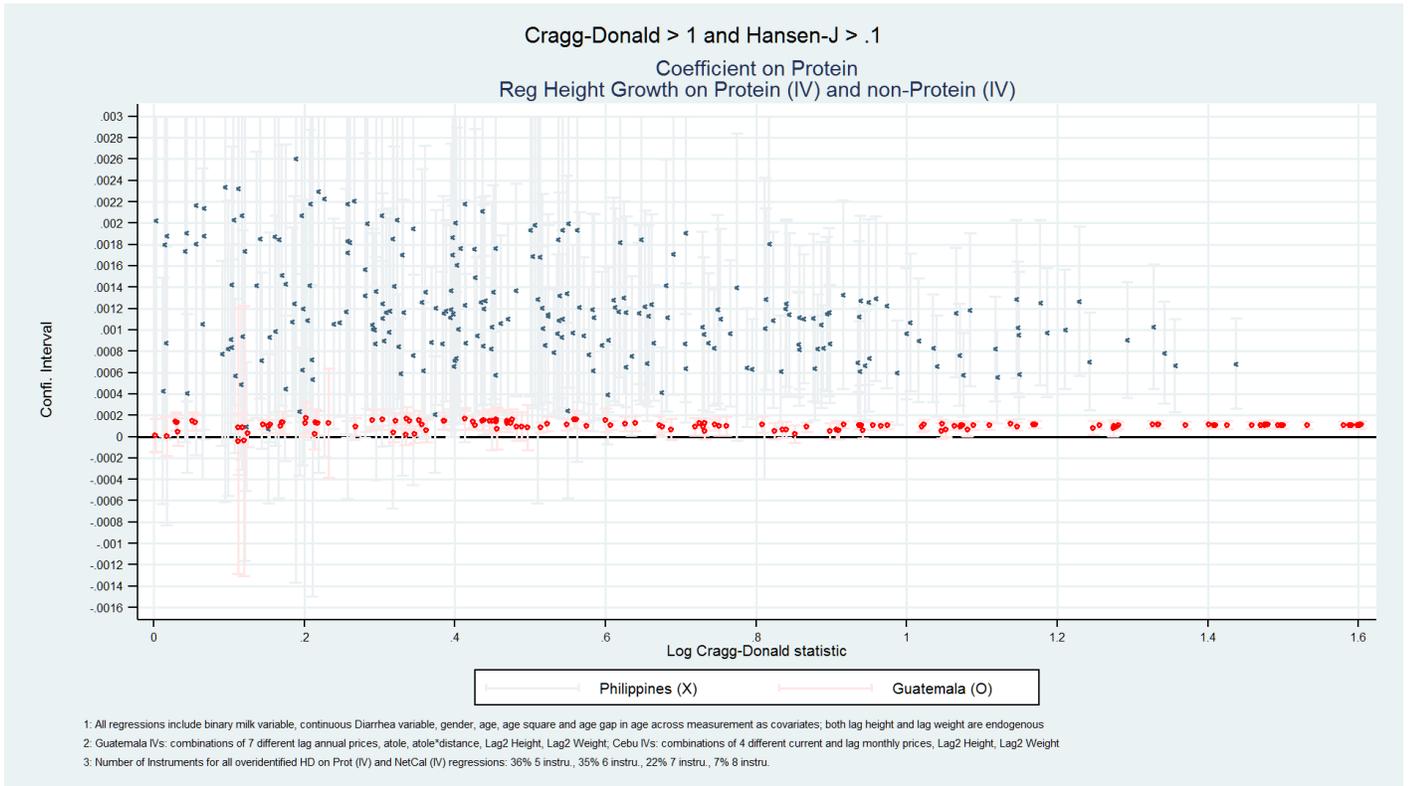

Change in Weight: Estimates $\lambda^w_{prot}$ from Equation (6) in the manuscript

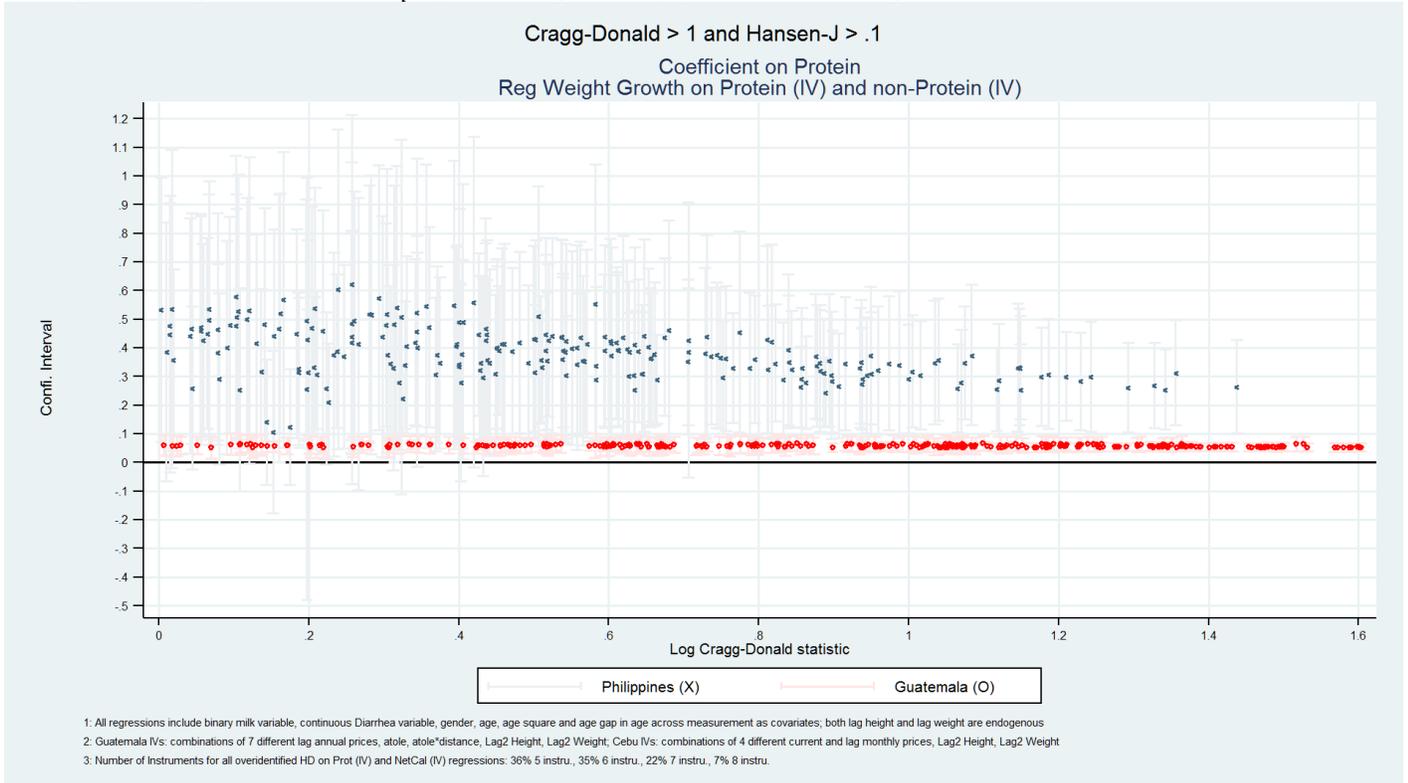

Figure A7: Non-Protein Coefficient
Change in Height: Estimates $\lambda^h_{non\_prot}$ from Equation (6) in the manuscript

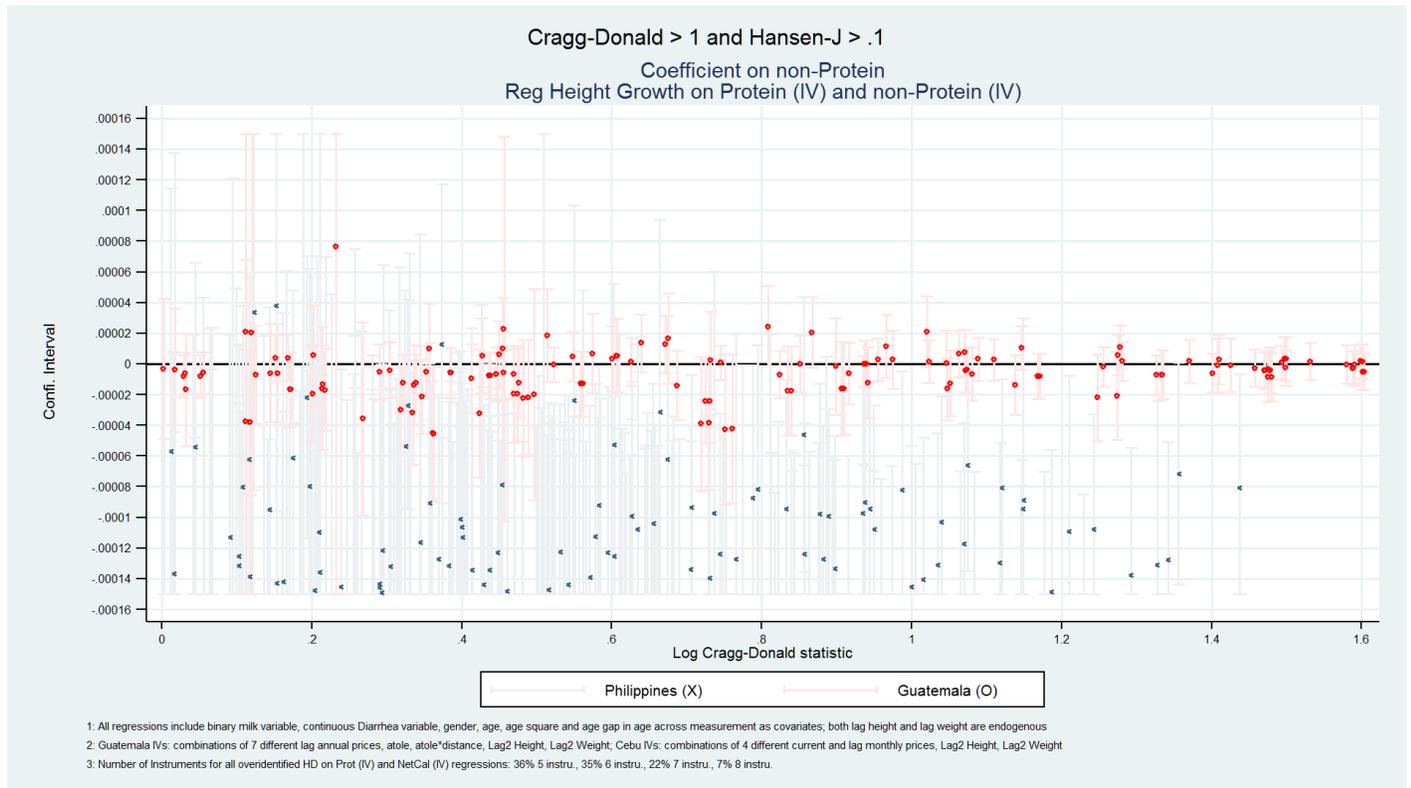

Change in Weight: Estimates $\lambda^w_{non\_prot}$ from Equation (6) in the manuscript

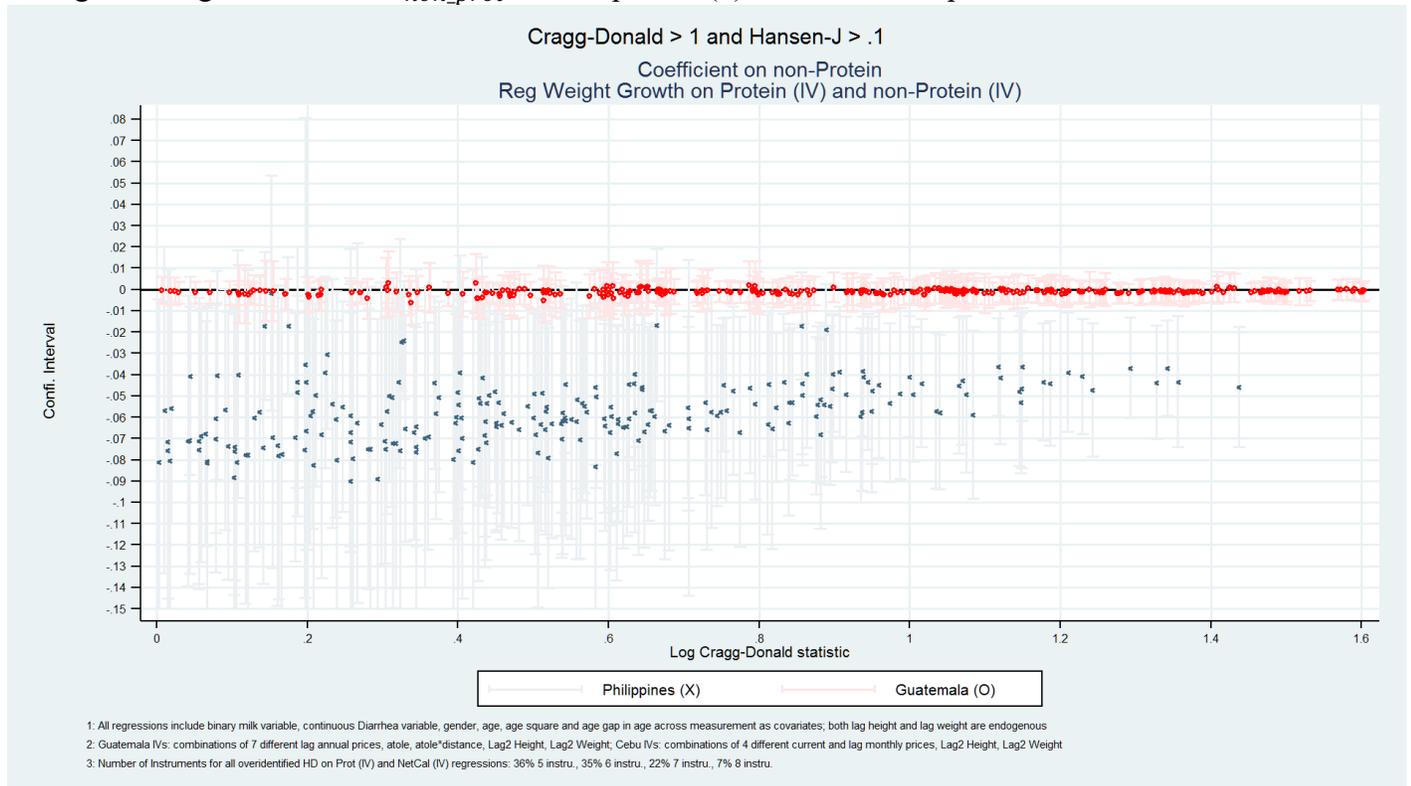

## 6. Specifications used in the tables and graphs

### Guatemala

Potential Instruments: Atole, Distance multiplied with Atole and prices of eggs, chicken, pork, beef, rice, beans, and corn.

Specifications used for Guatemala follow the following criteria:

1. Dependent variables: Change in Height and Change in Weight.
   Atole and all combinations of 3 or 4 of the 7 prices
Atole, Distance multiplied with Atole and all combinations of 2, 3 or 4 of the 7 prices

2. Dependent variable: Change in Weight.
Atole, second lag of height and all combinations of 2, 3 or 4 of the 7 prices
Atole, Distance multiplied with Atole, second lag of height and all combinations of 2, 3 or 4 of the 7 prices

3. Dependent variable: Change in Height
Atole, Distance multiplied with Atole, second lag of weight and all combinations of 2, 3 or 4 of the 7 prices
Atole, second lag of weight and all combinations of 2, 3 or 4 of the 7 prices

4. Dependent variables: Change in Height and Change in Weight
Atole, second lag of height and second lag of weight and all combinations of 2, 3 or 4 of the 7 prices
Atole, Distance multiplied with Atole, second lag of height and second lag of weight and all combinations of 2, 3 or 4 of the 7 prices

### Philippines

Potential Instruments: Second lag of height, second lag of height, current prices of tomatoes, corn, eggs, and dried fish, lagged prices of tomato, corn, egg and dried fish.

Specifications used for Philippines follow the following criteria:

1. Dependent Variables: Change in Height and Change in Weight
All combinations of 4, 5 or 6 of the 8 prices

2. Dependent variable: Change in weight
Second lag of height and all combinations of 3, 4, 5 or 6 of the 8 prices

3. Dependent variable: Change in Height:
Second lag of weigh and all combinations of 3, 4, 5 or 6 of the 8 prices

4. Dependent Variables: Change in Height and Change in Weight
Second lag of height and second lag of weight and all combinations of 2, 3, 4, 5 or 6 of the 8 prices